\documentclass[%
 aip,
 amsmath,amssymb,
 reprint,%
]{revtex4-1}

\usepackage{graphicx}
\usepackage{dcolumn}
\usepackage{bm}

\usepackage[utf8]{inputenc}
\usepackage[T1]{fontenc}
\usepackage{mathptmx}
\usepackage{etoolbox}
\usepackage{lineno}

\makeatletter
\let\MYcaption\@makecaption
\makeatother

\usepackage{subcaption}
\makeatletter
\let\@makecaption\MYcaption
\makeatother

\usepackage{comment}
\usepackage[colorlinks=true,linkcolor=blue,urlcolor=blue,citecolor=blue]{hyperref}

\newcommand\orcid[1]{\href{https://orcid.org/#1}{\includegraphics[height=9pt]{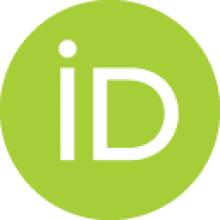}}}

\newcommand{\nocontentsline}[3]{}
\newcommand{\tocless}[2]{\bgroup\let\addcontentsline=\nocontentsline#1{#2}\egroup}

\begin{document}

\preprint{AIP/123-QED}

\title{The Simons Observatory: A fully remote controlled calibration system with a sparse wire grid for cosmic microwave background telescopes}

\author{Masaaki Murata~\orcid{0000-0003-4394-4645}}
\email{masaaki.murata@phys.s.u-tokyo.ac.jp}
\affiliation{Department of Physics, The University of Tokyo, Tokyo 113-0033, Japan\looseness=-1}

\author{Hironobu Nakata~\orcid{0000-0002-6300-1495}}
\affiliation{Department of Physics, Kyoto University, Kitashirakawa Oiwake-cho, Sakyo-ku, Kyoto 606-8502, Japan\looseness=-1}

\author{Kengo Iijima}
\affiliation{Department of Physics, The University of Tokyo, Tokyo 113-0033, Japan\looseness=-1}

\author{Shunsuke Adachi~\orcid{0000-0002-0400-7555}}
\affiliation{Department of Physics, Kyoto University, Kitashirakawa Oiwake-cho, Sakyo-ku, Kyoto 606-8502, Japan\looseness=-1}
\affiliation{Hakubi Center for Advanced Research, Kyoto University, Yoshida-honmachi, Sakyo-ku, Kyoto 606-850, Japan\looseness=-1}

\author{Yudai Seino~\orcid{0000-0001-5680-4989}}
\affiliation{Department of Physics, Kyoto University, Kitashirakawa Oiwake-cho, Sakyo-ku, Kyoto 606-8502, Japan\looseness=-1}
\affiliation{Joseph Henry Laboratories of Physics, Jadwin Hall, Princeton University, Princeton, NJ 08544, USA\looseness=-1}

\author{Kenji Kiuchi~\orcid{0000-0003-3243-8667}}
\affiliation{Department of Physics, The University of Tokyo, Tokyo 113-0033, Japan\looseness=-1}

\author{Frederick Matsuda~\orcid{0000-0003-0041-6447}}
\affiliation{Institute of Space and Astronautical Science, Japan Aerospace Exploration Agency (JAXA), Sagamihara, Kanagawa, 252-5210, Japan\looseness=-1}

\author{Michael J. Randall}
\affiliation{Department of Physics, University of California, San Diego, La Jolla, CA 92093, USA\looseness=-1}
\affiliation{Center for Astrophysics and Space Sciences, 9500 Gilman Dr, La Jolla, CA 92093, USA\looseness=-1}

\author{Kam Arnold}
\affiliation{Department of Physics, University of California, San Diego, La Jolla, CA 92093, USA\looseness=-1}
\affiliation{Center for Astrophysics and Space Sciences, 9500 Gilman Dr, La Jolla, CA 92093, USA\looseness=-1}

\author{Nicholas Galitzki~\orcid{0000-0001-7225-6679}}
\affiliation{Department of Physics, University of Texas at Austin, Austin, TX 78712, USA\looseness=-1}
\affiliation{Weinberg Institute for Theoretical Physics, Texas Center for Cosmology and Astroparticle Physics, Austin, TX 78712, USA\looseness=-1}

\author{Bradley R. Johnson~\orcid{0000-0002-6898-8938}}
\affiliation{Department of Astronomy, University of Virginia, Charlottesville, VA 22904, USA\looseness=-1}

\author{Brian Keating~\orcid{0000-0003-3118-5514}}
\affiliation{Department of Physics, University of California, San Diego, La Jolla, CA 92093, USA\looseness=-1}
\affiliation{Center for Astrophysics and Space Sciences, 9500 Gilman Dr, La Jolla, CA 92093, USA\looseness=-1}

\author{Akito Kusaka}
\affiliation{Department of Physics, The University of Tokyo, Tokyo 113-0033, Japan\looseness=-1}
\affiliation{Physics Division, Lawrence Berkeley National Laboratory, Berkeley, CA 94720, USA\looseness=-1}
\affiliation{Research Center for the Early Universe, School of Science, The University of Tokyo, Tokyo 113-0033, Japan\looseness=-1}
\affiliation{Kavli Institute for the Physics and Mathematics of the Universe (WPI), UTIAS, The University of Tokyo, Kashiwa, Chiba, 277-8583, Japan\looseness=-1}

\author{John B. Lloyd~\orcid{0000-0003-1581-1626}}
\affiliation{Department of Physics, University of California, San Diego, La Jolla, CA 92093, USA\looseness=-1}
\affiliation{Center for Astrophysics and Space Sciences, 9500 Gilman Dr, La Jolla, CA 92093, USA\looseness=-1}

\author{Joseph Seibert~\orcid{0000-0002-0298-9911}}
\affiliation{Department of Physics, University of California, San Diego, La Jolla, CA 92093, USA\looseness=-1}
\affiliation{Center for Astrophysics and Space Sciences, 9500 Gilman Dr, La Jolla, CA 92093, USA\looseness=-1}

\author{Maximiliano Silva-Feaver~\orcid{0000-0001-7480-4341}}
\affiliation{Department of Physics, University of California, San Diego, La Jolla, CA 92093, USA\looseness=-1}
\affiliation{Center for Astrophysics and Space Sciences, 9500 Gilman Dr, La Jolla, CA 92093, USA\looseness=-1}

\author{Osamu Tajima~\orcid{0000-0003-2439-2611}}
\affiliation{Department of Physics, Kyoto University, Kitashirakawa Oiwake-cho, Sakyo-ku, Kyoto 606-8502, Japan\looseness=-1}

\author{Tomoki Terasaki~\orcid{0000-0002-2380-0436}}
\affiliation{Department of Physics, The University of Tokyo, Tokyo 113-0033, Japan\looseness=-1}

\author{Kyohei Yamada~\orcid{0000-0003-0221-2130}}
\affiliation{Department of Physics, The University of Tokyo, Tokyo 113-0033, Japan\looseness=-1}


\begin{abstract}
For cosmic microwave background (CMB) polarization observations, calibration of detector polarization angles is essential. We have developed a fully remote controlled calibration system with a sparse wire grid that reflects linearly polarized light along the wire direction. The new feature is a remote-controlled system for regular calibration, which has not been possible in sparse wire grid calibrators in past experiments. The remote control can be achieved by two electric linear actuators that load or unload the sparse wire grid into a position centered on the optical axis of a telescope between the calibration time and CMB observation. Furthermore, the sparse wire grid can be rotated by a motor. A rotary encoder and a gravity sensor are installed on the sparse wire grid to monitor the wire direction. They allow us to achieve detector angle calibration with expected systematic error of $0.08^{\circ}$. The calibration system will be installed in small-aperture telescopes at Simons Observatory.

\end{abstract}

\maketitle


\section{Introduction}

Cosmic microwave background~(CMB) polarization is a powerful probe of cosmology.~\cite{Planck_cosmo_2020, Planck_infla_2020} CMB polarization can be decomposed into parity-even (curl-free) ``E-mode'' and parity-odd (divergent-free) ``B-mode'' patterns.
A typical model of inflation~\cite{Planck_infla_2020, Guth1981} predicts the B-mode induced by the primordial tensor perturbations propagating as gravitational waves.~\cite{Seljak1997} Therefore, primordial B-modes would be a key signature of inflation.
To perform an accurate observation of the B-mode,
calibration of detector polarization angles is essential.
Unintentionally rotating the angle over all of the detectors, i.e. mis-estimating ``the global polarization angle'' of detectors
will leak E-mode to B-mode power.~\cite{Keating2013,Kaufman2014}
Since the signal power of the E-mode is much larger than that of the B-mode,~\cite{Planck_cosmo_2020, Tristram2022} the leakage could bias a search for the primordial B-mode. 
In addition, angle calibration will allow us to use the EB cross-spectra for possibly detecting non-standard cosmology phenomena such as cosmic birefringence.~\cite{Carroll1998,Jost2023}

The power of the primordial B-mode signal can be characterized by the tensor-to-scalar ratio, $r$.
Simons Observatory (SO) aims to search for the primordial B-mode to a target sensitivity of $\sigma(r) = 0.003$.~\cite{Ade2019}
To achieve such a high sensitivity,
it is required that the systematic error of the global polarization angle is within 
$\sim 0.1^{\circ}$ because the angle miscalibration introduces a non-negligible bias on the measurement of $r$.~\cite{Bryan2018,Abitbol2021,Johnson2015}

Previous CMB experiments~\cite{Takahashi2008, Polarbear2014, Bischoff2011} have calibrated the polarization angle with an artificial calibrator, such as polarizing dielectric sheets, or an observation of a polarized astronomical source, such as the Moon or Tau A. Polarizing dielectric sheets reflect or transmit incident light and produce polarized light.~\cite{ODell2002} The BICEP1 experiment calibrated polarization angles of detectors with $\Delta\theta < 0.7^{\circ}$ setting a polarization dielectric sheet in front of the telescope.~\cite{Takahashi2008} Tau A is a polarized supernova remnant and the characteristics have been measured by several experiments in the frequency range from 23 to 353 GHz.~\cite{Weiland2011, Aumont2010, Planck2016, Aumont2020} The \textsc{Polarbear} experiment calibrated polarization angles of detectors with $\Delta\theta < 0.43^{\circ}$ observing Tau A.~\cite{Polarbear2014} Thermal radiation from the Moon is polarized normal to the lunar surface.~\cite{Bischoff2008} The QUIET experiment calibrated polarization angles of detectors with $\Delta\theta < 3^{\circ}$ observing the Moon.~\cite{Bischoff2011} 

The conventional calibration methods have several limitations. In the artificial calibrator case, it requires time and personnel to fix the calibrator to the telescope for the calibration time and also to remove it for the CMB observation. This requirement makes it difficult to perform the calibration frequently. In the case of polarized astronomical sources, the time to see the source is limited. However, the regular angle calibration is important to measure the time-variant cosmic birefringence which is predicted in the context of particle physics beyond the standard model like axion-like particles.~\cite{Fedderke2019} Therefore, a new calibrator that can calibrate the polarization angles of detectors remotely and regularly is essential.

To calibrate the angle remotely and regularly, we have developed a new angle calibration system.
The calibration system will be installed into the 42-cm small-aperture telescopes
(SATs) of SO.~\cite{Galitzki2018,Ali2020,Kiuchi2020,Galitzki2023}
SO will deploy one 6-m large-aperture telescope~\cite{Parshley2018,Zhu2021} and an initial set of three SATs followed by another three SATs supported by SO:UK and SO:Japan at altitude of $5200\mathrm{\,m}$ in the Atacama Desert of Chile.
The SATs are optimized to detect the primordial B-mode and the initial three SATs will together contain $>30000$~transition-edge-sensor (TES) detectors.~\cite{Irwin2005}
Observations will be made at six frequency bands: 27/39 GHz (low frequency (LF)), 93/145 GHz (mid-frequency (MF)), and 225/280 GHz (ultra-high frequency (UHF)). 
Each TES detector on a focal plane is coupled to a sinuous antenna with a lenslet at LF or an orthomode transducer (OMT) with a horn at MF and UHF.~\cite{McCarrick2021} The sinuous antenna and the OMT receive an incoming polarization signal and send it to the TES detector. We calibrate the polarization angles of them.
To suppress the 1/f noise, each SAT will have a continuously rotating cryogenic half-wave plate (CHWP),~\cite{Kusaka2014,Takakura2017,Yamada2022} which modulates incoming polarization signals.
In addition to our calibration system, angle calibrations using Tau A and an artificial source with a drone~\cite{Coppi2022} are planned.
We will compare our calibration system with the drone and Tau A to check the consistency.~\cite{Abitbol2021}
This paper describes the concept of the calibration system (Section~\ref{sec:cal}), the requirements for the system (Section~\ref{sec:req}), and the design and evaluation to confirm that the system satisfies the requirements (Sections~\ref{sec:design}, and \ref{sec:meas}).

\section{Concept}\label{sec:cal}

\begin{figure}
\centering\includegraphics[width=\linewidth]{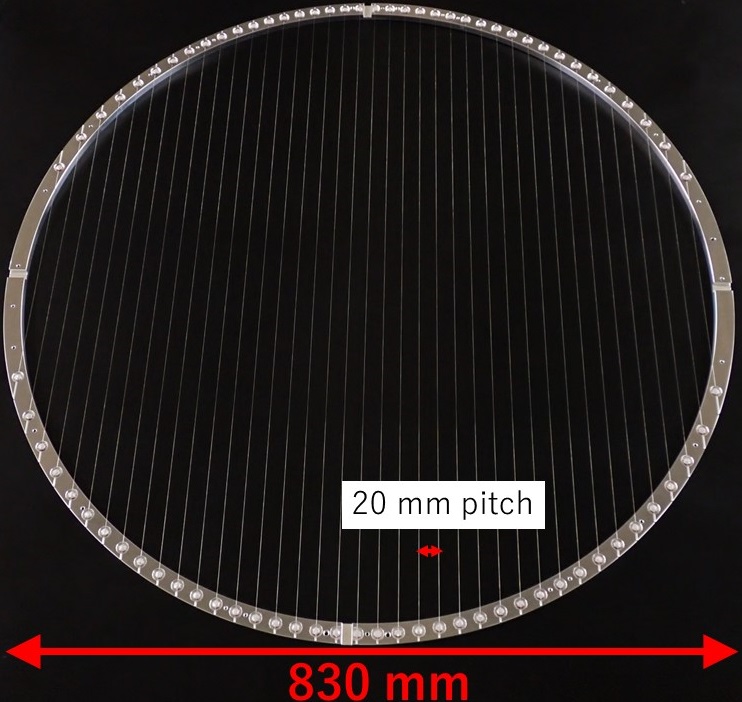}
\caption{Sparse wire grid (SWG). This SWG is for the SAT MF band and has 39 wires stretched on an aluminum ring with 20-mm pitch.}
\label{fig:gridring}
\end{figure}

The new calibration system is based on one of the artificial light sources, a sparse wire grid (SWG) as shown in Fig.~\ref{fig:gridring}. 
The SWG is composed of parallel metal wires with a sparse pitch that is much larger than the wavelength of the observed light. The SWG reflects $300\mathrm{\,K}$ blackbody radiation from surrounding materials. The reflected light has a broadband spectrum and is polarized along the wires.
The SWG has another advantage that it is an aperture-filling source and calibrates all detectors on a focal plane at once with a large optical throughput, i.e. large aperture and large solid angle, while maintaining manageable size. This feature enables accurate calibration of the relative angles among all of the detectors in addition to the global angle calibration.

The SWG can calibrate not only the polarization angle but also responsivity, a detector time constant and polarization efficiency.
We determine time-dependent relative responsivity between detectors.~\cite{Tajima2012} 
By combining the CHWP and the SWG, we can make accurate in-situ optical time constant measurements using the phase delay in the CHWP signal modulation when varying the rotation speed of the CHWP.~\cite{Simon2014,Kusaka2018}
In addition, we determine a polarization efficiency of detectors~\cite{Simon2014} independently of the modulation efficiency of CHWP.

The calibration system has a new feature of a fully remote-controlled system to move the SWG for regular calibration. The SWG is placed in the optical path of the telescope for calibration but must be removed from the optical path during CMB observation to avoid any optical influence. The past experiments, the QUIET~\cite{Tajima2012} and the Atacama B-mode Search (ABS),~\cite{Kusaka2018} used SWGs as calibrators but had no system to automatically load the SWG into a position centered on the optical axis. 
The QUIET and the ABS conducted the calibration once a season~\cite{Chinone2010} and several times a season~\cite{Kusaka2018} respectively.
Our calibration system can load and unload the SWG into a position centered on the optical axis using linear actuators so that we can perform the calibration remotely and regularly. This calibration is planned to be run a few times per day or more.
This regular calibration of our calibration system will allow us to calibrate the time trend of the relative angles for each detector.

\begin{figure}
    \begin{minipage}{0.48\textwidth}%
        \centering
        \includegraphics[height=5.5cm]{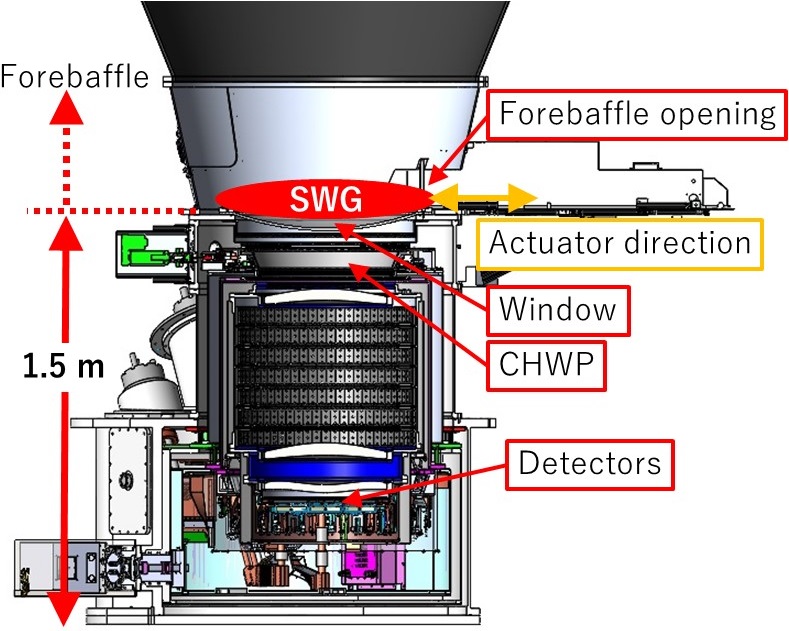}%
        \subcaption{}
        \label{fig:SAT_cross_section_a}
    \end{minipage}%
    \\
    \begin{minipage}{0.48\textwidth}
        \centering
        \includegraphics[height=5cm]{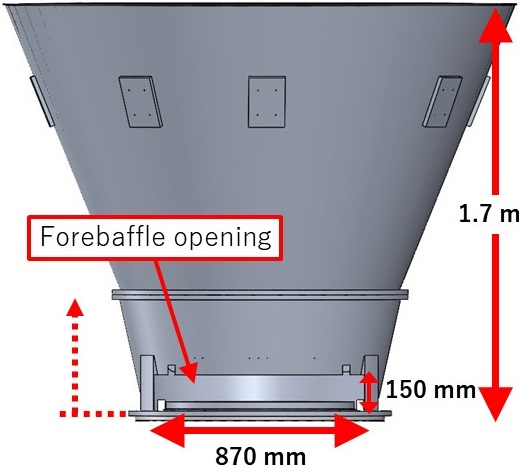}%
        \subcaption{}
        \label{fig:SAT_cross_section_b}
    \end{minipage}
    \caption{
    (\subref{fig:SAT_cross_section_a}) Cross section of the SAT receiver, which also shows the SWG position during the calibration, and (\subref{fig:SAT_cross_section_b}) overview of forebaffle from the right side view of (\subref{fig:SAT_cross_section_a}). The SWG is positioned above the window inside the forebaffle and moves from the inside of the forebaffle through the forebaffle opening using linear actuators.
    }
    \label{fig:SAT_cross_section}
\end{figure}

The SAT receiver is a cryogenic three-lens refractor. A large baffle (a ``forebaffle'' with a $2.08$-m-diameter upper aperture) is installed above the cryostat window on the SAT to suppress stray light from the nearby terrain. The SWG is positioned $70\mathrm{\,mm}$ above the window inside the forebaffle to illuminate the detectors in the cryostat (Fig.~\ref{fig:SAT_cross_section_a}). To move the SWG out of the optical path after the calibration, the forebaffle has been designed with an opening (Fig.~\ref{fig:SAT_cross_section_b}). We have designed a system using linear actuators to remove the SWG from the inside of the forebaffle through the opening.




\begin{figure}
\centering\includegraphics[height=6.5cm]{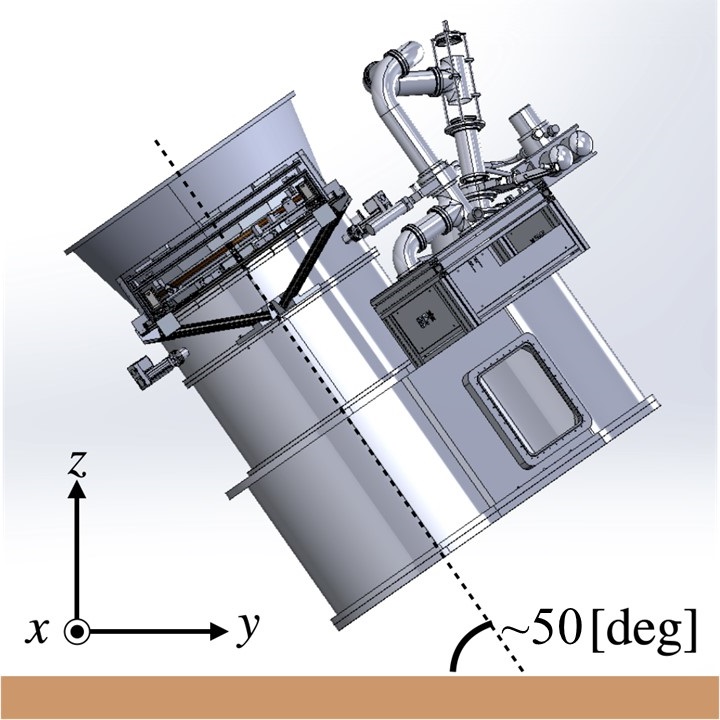}
\caption{Orientation of the telescope and SWG during the angle calibration. The SAT will be set at an elevation around $50^{\circ}$ from the horizontal plane for observation. The coordinate system is fixed to the ground. In this figure, the top section of the forebaffle is not displayed.}
\label{fig:wiregrid_position}
\end{figure}

The calibration system also has a remote-controlled system for the SWG rotation. In the calibration, we plan to rotate the SWG with steps of $22.5^{\circ}$ and perform the calibration of detector polarization angles at each SWG angle to understand the influence on the calibration caused by the wire angle difference. The rotation is enabled by a motor. A rotary encoder and a gravity sensor are installed in the system to measure the rotation angle and the alignment of the rotation plane, respectively. They allow us to know the accurate direction of the wires and to achieve a global angle measurement with a systematic error of $0.08^{\circ}$. The evaluation of the systematic error is described later in Section~\ref{sec:meas}.
The remote-controlled system with the linear actuators and the motor significantly improves the repeatability in the grid placement and rotation compared to the SWGs in the past experiments.

\begin{figure}
    \begin{minipage}{0.48\textwidth}%
        \centering
        \includegraphics[height=5.5cm]{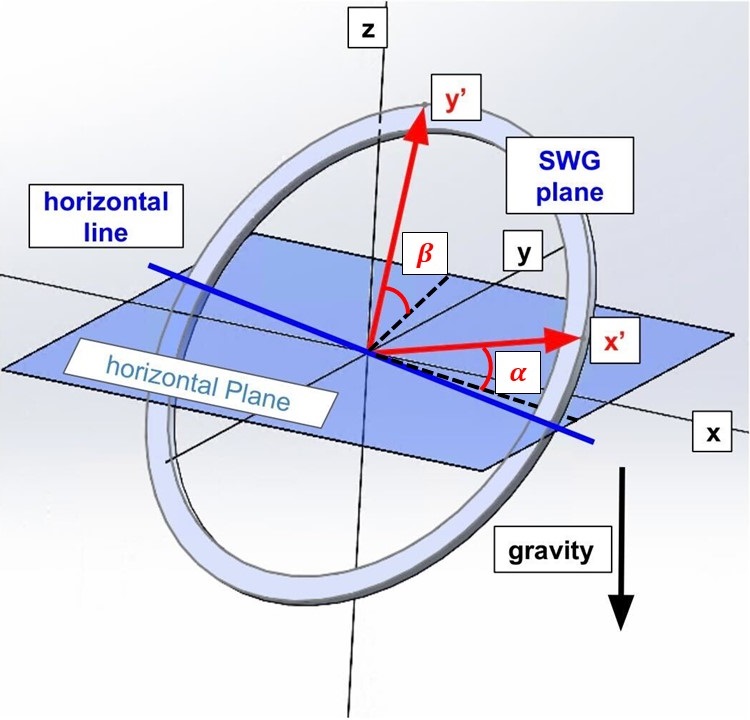}%
        \subcaption{}
        \label{fig:gravitytest_a}
    \end{minipage}%
    \\
    \begin{minipage}{0.48\textwidth}
        \centering
        \includegraphics[height=5cm]{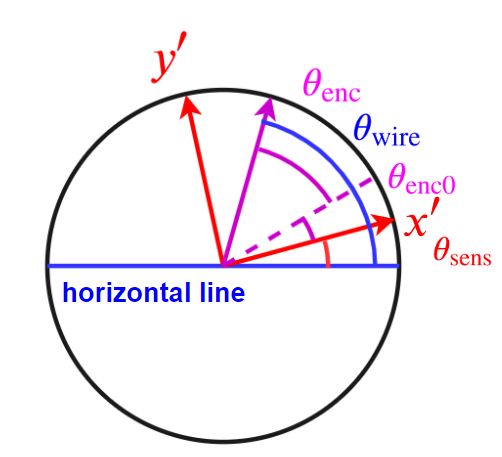}%
        \subcaption{}
        \label{fig:gravitytest_b}
    \end{minipage}
    \caption{
    (\subref{fig:gravitytest_a}) Orientation of the SWG plane in a $xyz$ coordinate system fixed to the ground and (\subref{fig:gravitytest_b}) definitions of the angle parameters in the SWG plane. The negative $z$ axis represents the direction of gravity. The rectangle in (\subref{fig:gravitytest_a}) represents the horizontal plane ($xy$ plane). The SWG plane is the same plane as the $x^{\prime}y^{\prime}$ plane defined by the two-axis gravity sensor. The horizontal line (blue line) is shared by the SWG plane and horizontal plane. $\theta_{\mathrm{wire}}$ is the angle between the wire direction and the horizontal line. The angle between the horizontal line and the $x^{\prime}$ axis is defined as $\theta_{\mathrm{sens}}$. $\theta_{\mathrm{enc}}$ is an angle measured with an encoder and $\theta_{\mathrm{enc}0}$ is its offset.
    }
    \label{gravitytest}
\end{figure}

The SAT including the calibration system is mounted on the 3-axis-actuation SAT platform~\cite{Ali2020, Galitzki2023} and can change the azimuth and elevation of the pointing direction and rotate around the optical axis.
We plan to nominally set the SAT at an elevation around $50^{\circ}$ from the horizontal plane for observation.~\cite{Kiuchi2020,Galitzki2023} 
The calibration will be performed at the nominal elevation shown in Fig.~\ref{fig:wiregrid_position}.
Since the polarization angle of the light from the SWG is parallel to the wires,
measuring the wire orientation in the SWG plane is essential for this calibration.
In particular, to project the wire orientation on the sky map, it is important to measure the rotation angle of the wires $\theta_\mathrm{wire}$ in the SWG plane from a ``horizontal line'' where the SWG plane and the horizontal plane intersect each other (Fig.~\ref{gravitytest}).

To calculate the angle $\theta_\mathrm{wire}$ between the wires and horizontal line, we measure four angles: $\alpha$, $\beta$, $\theta_{\mathrm{enc}}$  and $\theta_{\mathrm{enc}0}$.
$\alpha$ and $\beta$ are the angles between the horizontal plane and SWG plane, which are measured by a two-axis gravity sensor installed on the SWG plane and the measured axes ($x^{\prime}$ and $y^{\prime}$) on the SWG plane are defined by the gravity sensor.
The gravity sensor is roughly aligned to make the $x^{\prime}$ axis parallel to the horizontal line.
$\alpha$ and $\beta$ are also measured by encoders of the SAT platform and the measured values can be used as a cross-check toward the gravity sensor.
$\theta_{\mathrm{enc}}$ is the rotation angle of the wires measured by the encoder.
$\theta_{\mathrm{enc}0}$ is the offset of the encoder output. 
By combining them, $\theta_\mathrm{wire}$ can be calculated as
\begin{equation}\label{eq:angle_calc}
    \theta_{\mathrm{wire}} =\theta_{\mathrm{enc}} + \theta_{\mathrm{enc}0} +  \theta_{\mathrm{sens}},
\end{equation}
where
\begin{equation}
    \theta_{\mathrm{sens}} =  \arctan{\left( \frac{\sin{\alpha}}{\sin{\beta}} \right)}.
\end{equation}
$\theta_{\mathrm{enc}0}$ is defined as the difference between $\theta_{\mathrm{enc}}$ and $\theta_{\mathrm{wire}}$ when $x^{\prime}$ is aligned along the horizontal line ($\alpha=0$). Since $\theta_{\mathrm{enc}0}$ is time-invariant and independent of $\theta_{\mathrm{wire}}$, $\theta_{\mathrm{enc}0}$ is measured in a laboratory using an extra gravity sensor as described in Section~\ref{sec:meas}.
$\theta_{\mathrm{sens}}$ is a small rotation angle of the $x^{\prime}$ axis with respect to the horizontal line around the optical axis. It is ideally zero if the gravity sensor is exactly aligned to the horizontal plane. 
The systematic error on $\theta_{\mathrm{wire}}$ calculated in Eq.~(\ref{eq:angle_calc}) using the encoder and gravity sensor must be within $0.1^{\circ}$ to achieve the SO science target of measuring the tensor-to-scalar ratio with $\sigma(r)=0.003$ 
using solely the calibration system to calibrate the polarization angle.

\section{Requirements}\label{sec:req}
The calibration system has requirements on the sizes and the actuators in addition to the requirement on the angle systematic error. Table~\ref{tab:requirement} summarizes these requirements.

\begin{table}[h!]
\begin{center}
\caption{Key system dimensions and requirements for the calibration system}
\label{tab:requirement}
\begin{tabular}{|c|c|c|} \hline
Parameter & Requirement \\ \hline
Wire angle systematic error & $<0.1^{\circ}$\\
SWG diameter & $>720\ \rm{mm}$\\
Width of calibration system & $<870\ \rm{mm}$\\
Height of calibration system & $<150\ \rm{mm}$\\
Actuator length & $>1.5\ \rm{m}$\\ 
Actuator loadable weight & $>23\,\mathrm{kg}$\\
Actuator insertion time & $\lessapprox3\,\mathrm{minutes}$\\ 
Actuator positioning accuracy & $\lessapprox1\,\mathrm{mm}$\\ 
Rotation positioning accuracy & $\lessapprox1^{\circ}$\\
Calibration duration & $\lessapprox10\,\mathrm{minutes}$\\
 \hline
\end{tabular}
\end{center}
\end{table}

\begin{figure}
\centering\includegraphics[width=7cm]{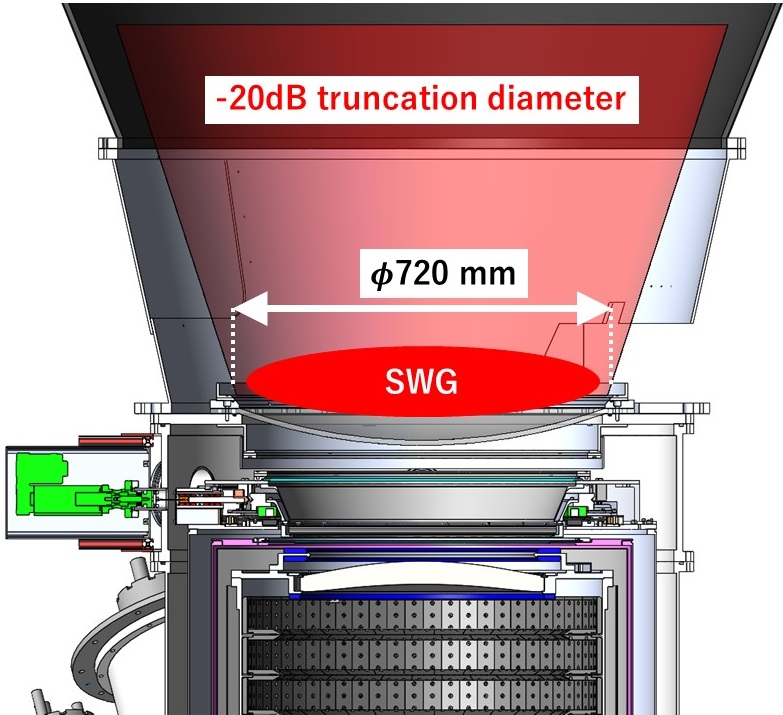}
\caption{Cross section of the SAT receiver around the installation position of the SWG and the diameter of the beam defined by $-20\,\mathrm{dB}$ truncation.}
\label{fig:-20dBcone}
\end{figure}

The diameter of the beam is defined by a $-20\,\mathrm{dB}$ truncation diameter, which designates the region where diffraction of the beam of edge detectors on the focal plane becomes $-20\,\mathrm{dB}$ assuming the uniform illumination at the aperture.~\cite{Matsuda2022}
The SWG installation position requires an inner diameter greater than $720\,\mathrm{mm}$ to accommodate the $-20\,\mathrm{dB}$ truncation diameter of the beam~(Fig.~\ref{fig:-20dBcone}).
On the other hand, the SWG size can not be arbitrarily large as it must fit through the designated $870\,\mathrm{mm} \times 150 \,\mathrm{mm}$ opening in the forebaffle.
Automatic switching of the SWG position is enabled by two electric linear actuators.
The actuators also have several requirements.
To move the whole of the SWG out of the forebaffle, the actuator length must be more than 1.5 m.
Since the telescope changes the elevation and rotates around the optical axis, the actuators need to carry the SWG at any inclination.
Therefore, the loadable weight of the two actuators in moving vertically must be more than $23\,{\mathrm{kg}}$. This weight corresponds to the sum of the SWG assembly weight and an additional load to move the cover system, which is described in Section~\ref{sec:design}.
We also set other requirements as shown in Table~\ref{tab:requirement} for the actuator insertion time, the actuator positioning accuracy, the rotation positioning accuracy, and the operation time.
The actuator insertion time is the time of loading or unloading the SWG assembly into a position centered on the optical axis using linear actuators.
The actuator positioning accuracy is the accuracy for placing the SWG assembly in the optical path of the telescope.
The rotation positioning accuracy is the accuracy for rotating the SWG with the step of $22.5^{\circ}$.
The calibration duration is the time taken for all of the operations: the insertion of the SWG assembly into the forebaffle, 16 rotations of the SWG by $22.5^{\circ}$ each, and the removal.

\section{Design}\label{sec:design}



\begin{figure}
\centering\includegraphics[width=7cm]{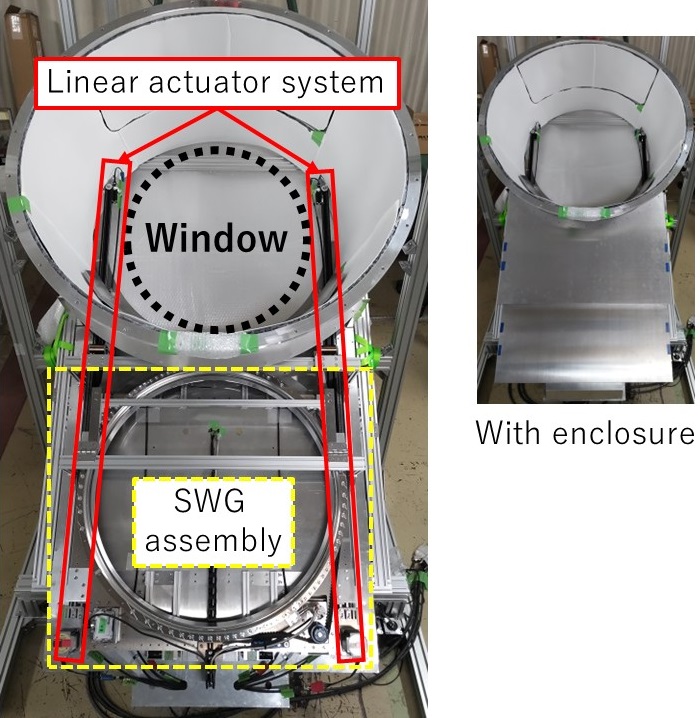}
\caption{Overview of the calibration system with the forebaffle bottom section. The calibration system consists of the SWG and traveling system with two linear actuators as shown in the left photo. The calibration system is covered with an aluminum enclosure as shown in the right photo.}
\label{fig:overview_photo}
\end{figure}

Figure~\ref{fig:overview_photo} shows an overview of the calibration system.
The calibration system consists of an SWG assembly and linear actuator system.
The SWG assembly has a baseplate on which a rotation bearing is mounted with the SWG. 
The baseplate is fastened to the movable plates of the two linear actuators.
The calibration system is covered with an aluminum enclosure to prevent dust, rain, and ambient light entering the inside of the forebaffle through the forebaffle opening.
The inside of the enclosure is blackened with a microwave absorber (E$\&$C Engineering AN-W72~\cite{EC}).


\subsection{SWG}
The SWG for the SAT MF band has 39 wires stretched on a ring with 20-mm pitch.
The amplitude of the polarization signal is proportional to the products of the inverse of the wire pitch, beam illumination on the SWG, and the difference between the reflected ambient temperature and the background load temperature.~\cite{Tajima2012}
In the ABS experiment,~\cite{Kusaka2014} the polarization signal of the SWG was $\sim0.5\,\mathrm{K}$. 
Considering the differences in the wire pitch and beam size between that SWG and ours, the polarization signal in our system is expected to be $\mathcal{O}(1)\mathrm{\,K}$, which is much higher than the white noise level of the SAT detectors.~\cite{Ade2019}
The wire is made of tungsten to mitigate creep in the wire and the diameter is $0.1\mathrm{\,mm}$.
The diameter is sufficiently smaller than the observation wavelength,
which is $\gtrsim1\mathrm{\,mm}$.
The cross polarization by reflection of the SWG is smaller than $1\%$ of the polarization signal along the wires considering the wire diameter and the observation wavelength (Appendix~\ref{sec:cross_pol}). The skin depth of tungsten is $\sim 0.4\ \mu \mbox{m}$ in the SAT MF band, which can be calculated from the tungsten resistivity of $5 \times 10^{-8} \Omega\mathrm{\,m}$ at $300\mathrm{\,K}$.~\cite{Weast2011} These properties lead to the SWG's reflectivity of $\sim 1$.  Thus, the light from SWG is primarily due to reflection, not emission.

\begin{figure}
\centering\includegraphics[width=7cm]{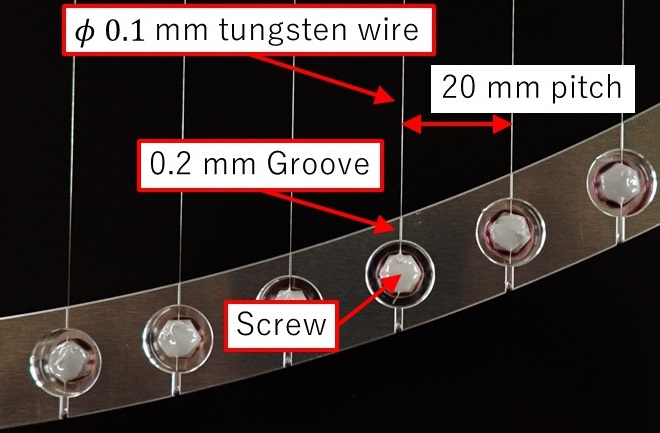}
\caption{Grooves and screws on the aluminum ring. The wires are glued on the heads of screws fixed on the aluminum ring.}
\label{fig:wiregroove}
\end{figure}

In the SWG, the wires are epoxied to screws mounted on an aluminum ring.
The ring has thin grooves and tapped holes arranged with the grooves (Fig.~\ref{fig:wiregroove}).
The wires are hooked over the grooves and are glued on the heads of screws fixed in the tapped holes. Broken wires can be replaced by unscrewing the screws without any damage to the ring.
The wire alignment is set by the machined grooves, each of which is only $0.2\,\mathrm{mm}$ wide.
While the adhesive cures, the wires are tensioned by $230$-g weights to minimize sagging.  
At this tension of $290\,\mbox{N}/\mbox{mm}^2$, the expected sag is only $0.05\,\mbox{mm}$ across the $743\,\mbox{mm}$ length, which corresponds to $\Delta\theta < 0.01^{\circ}$. We describe measurements of the sag in Section~\ref{sec:meas}. 
This tension is relatively small compared to tungsten's tensile strength of $2940\,\mbox{N}/\mbox{mm}^2$, so does not lead to any creep in the wires.~\cite{Rieck1967}


\subsection{SWG assembly}
\begin{figure}
     \begin{minipage}{0.48\textwidth}%
        \centering
        \includegraphics[width=7cm]{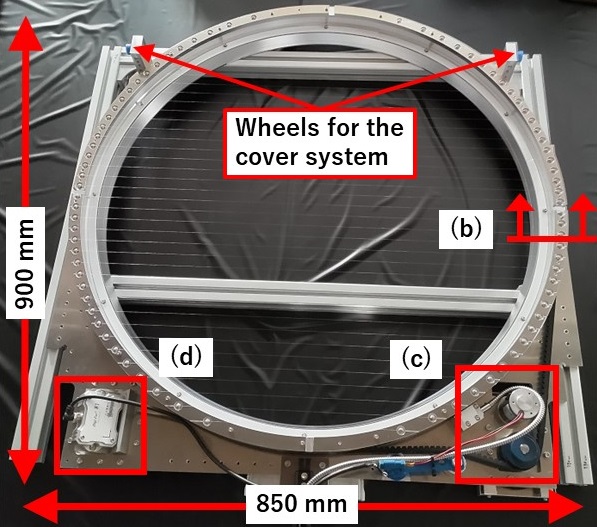}%
        \subcaption{}
        \label{fig:wire_grid_assembly}
    \end{minipage}%
    \\
    \begin{minipage}{0.48\textwidth}%
        \centering
        \includegraphics[height=4.5cm]{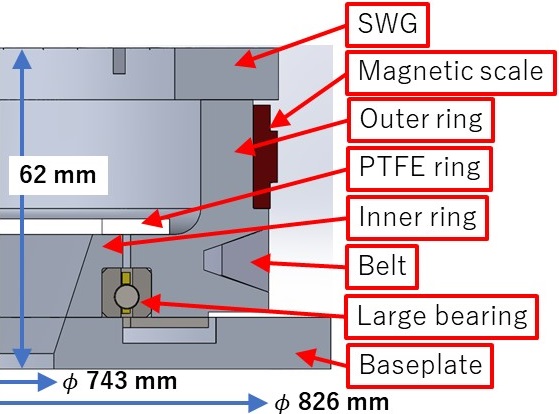}%
        \subcaption{}
        \label{fig:bearing}
    \end{minipage}%
    \\
    \begin{minipage}{0.48\textwidth}
        \centering
        \includegraphics[height=4.5cm]{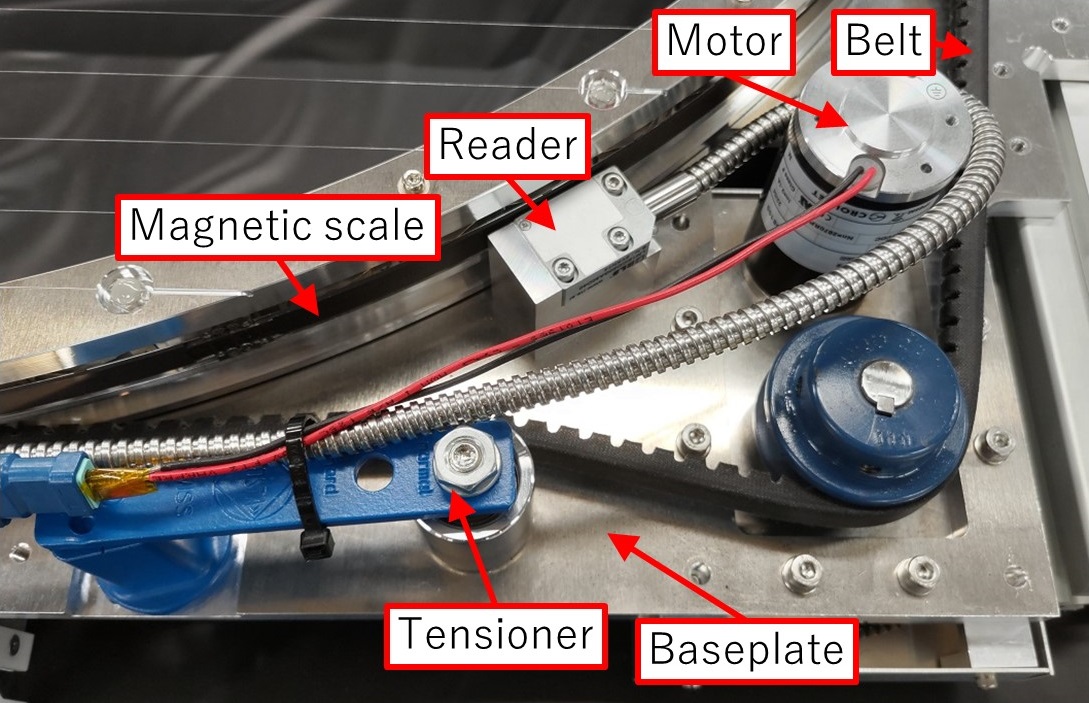}%
        \subcaption{}
        \label{fig:baseplate_parts}
    \end{minipage}
    \\
    \begin{minipage}{0.48\textwidth}
        \centering
        \includegraphics[height=4.5cm]{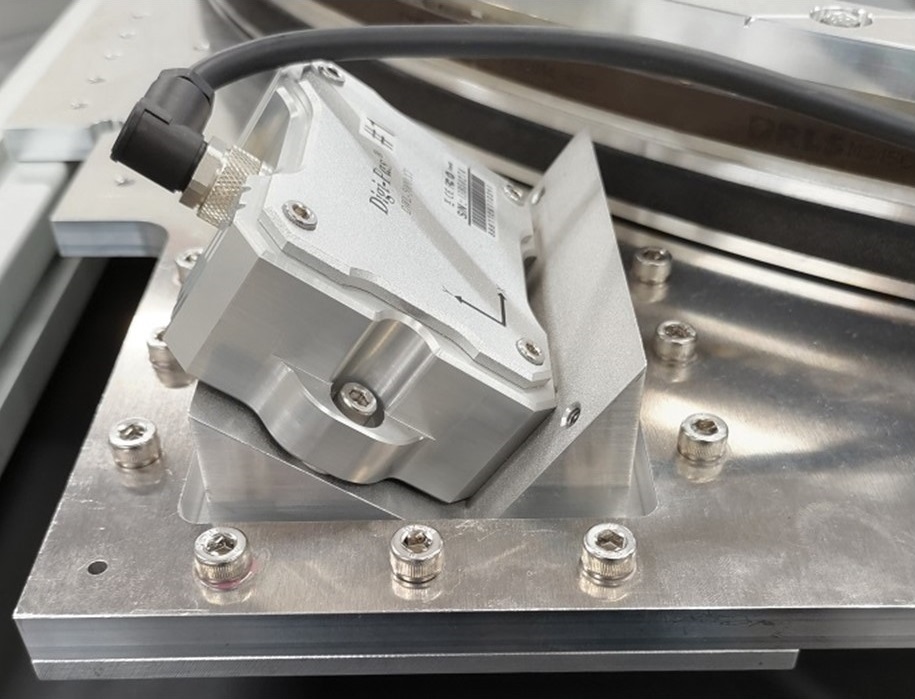}%
        \subcaption{}
        \label{fig:gravitysensor}
    \end{minipage}
    \caption{
    (\subref{fig:wire_grid_assembly}) SWG assembly,
    (\subref{fig:bearing}) cross section of the rotation bearing of the SWG assembly, 
    (\subref{fig:baseplate_parts}) mechanical parts installed on the baseplate for rotating the SWG, and
    (\subref{fig:gravitysensor}) gravity sensor.
    }
    \label{fig:wire_grid_overview}
\end{figure}

Figure~\ref{fig:wire_grid_assembly} shows an overview of the SWG assembly. The SWG is rotated on a large bearing (Silverthin SC300XP0,~\cite{Silverthin} $\phi 762\,\mathrm{mm}$). 
The inner ring of the bearing is fastened on the baseplate as a stator.
The outer ring fastens the SWG, as shown in Fig.~\ref{fig:bearing}.
The height of the SWG assembly is $62\,\mathrm{mm}$, and the height including the linear actuator system is $107\,\mathrm{mm}$.  The SWG’s innermost and outermost diameters are $743\,\mathrm{mm}$ and $826\,\mathrm{mm}$, respectively.  These dimensions satisfy the requirements in Table~\ref{tab:requirement}.
The gap of the inner and outer rings of the bearing is covered with a PTFE ring for weather and dust proofing.
A belt passing through the belt groove on the outer ring is driven by a motor on the baseplate~(Fig.~\ref{fig:baseplate_parts}). It rotates the outer ring with the SWG. The belt tension is maintained by a tensioner installed on the baseplate.

\subsection{Encoder}
The encoder is an incremental magnetic encoder (Renishaw LM15 IC~\cite{Renishaw}) and consists of a magnetic scale and a reader.
The magnetic scale is wrapped along the outer ring, and the reader is set on the baseplate.
When the outer ring rotates, the reader generates signals representing the counts of the magnetic scale and the rotation direction.
The magnetic scale has $52000$ counts per lap, which corresponds to an angle resolution of $0.002^{\circ}$.
The magnetic scale has one reference point in a lap, 
at which the encoder angle $\theta_{\mathrm{enc}}$ is $0^{\circ}$.


\subsection{Gravity sensor}
A gravity sensor is installed on the baseplate to monitor the orientation of the SWG plane, as shown in Fig.~\ref{fig:gravitysensor}.
The gravity sensor is a Digi-pas DWL-5000XY,~\cite{Digi-pas} which can measure elevation angles on two axes simultaneously. The angle accuracy according to the sensor’s specifications is $0.01^{\circ}$ at $0^{\circ}$ to $\pm 2.00^{\circ}$ and $0.03^{\circ}$ at other angles for both axes. The gravity sensor is tilted by $40^{\circ}$ toward the SWG plane to set the measured angle around $0^{\circ}$  at the telescope posture during the calibration, as shown in Fig.~\ref{fig:wiregrid_position}. Thus, the measurement accuracies of $\alpha$ and $\beta$ are $ 0.01^{\circ}$.



\subsection{Linear actuators}
\begin{figure}
    \begin{minipage}{0.48\textwidth}%
        \centering
        \includegraphics[height=3cm]{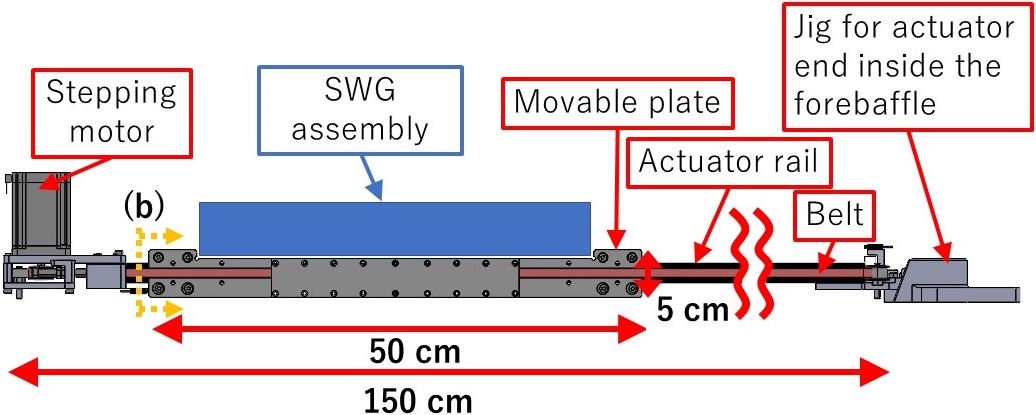}%
        \subcaption{}
        \label{fig:actuator_overview}
    \end{minipage}
    \\
    \begin{minipage}{0.48\textwidth}
        \centering
        \includegraphics[height=5cm]{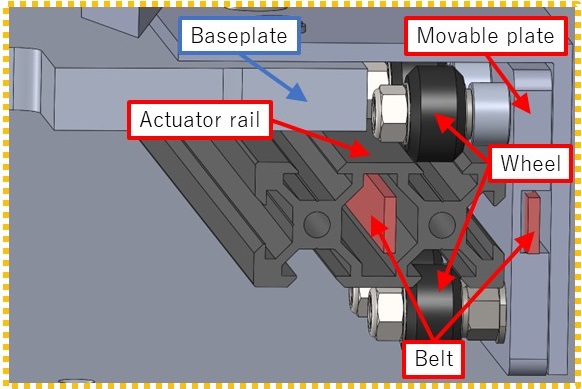}%
        \subcaption{}
        \label{fig:actuator_cross_section}
    \end{minipage}
    \caption{
    (\subref{fig:actuator_overview}) Overview of the actuator, (\subref{fig:actuator_cross_section}) cross section of the actuator from the left side view of (\subref{fig:actuator_overview}).
    }
    \label{fig:actuator}
\end{figure}

We used two 1.5-m-long linear actuators (Openbuilds V-slot NEMA 23 Linear Actuator~\cite{OpenBuilds}) with stepper motors (Orientalmotor PK269JDA~\cite{Orientalmotor}). 
The stepper motors rotate belts passing through the actuator rails as shown in Fig.~\ref{fig:actuator}.
The baseplate is fastened to two movable plates on the actuators, each of which has 8 wheels moving on the rails.
At the ends of the actuator rails, limit switches (Omron D2JW-01K31-MD~\cite{Omron}) are installed and can position the baseplate with an accuracy of $0.8\,\mathrm{mm}$.

\subsection{Cover system}
\begin{figure}
    \begin{minipage}{0.48\textwidth}%
        \centering
        \includegraphics[height=3.5cm]{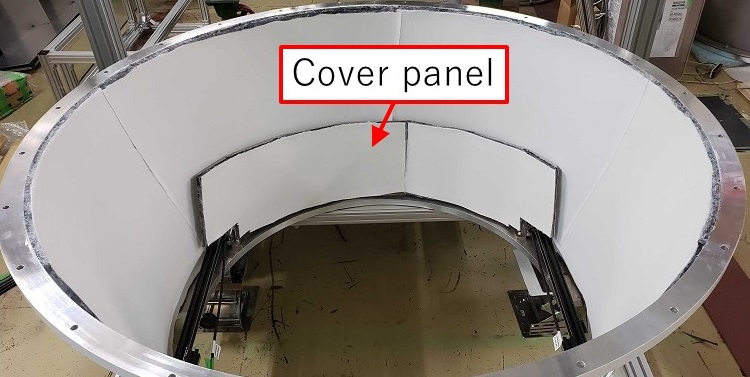}
        \subcaption{}
        \label{fig:shutter_overview}
    \end{minipage}
    \\
    \begin{minipage}{0.48\textwidth}
        \centering
        \includegraphics[height=2.5cm]{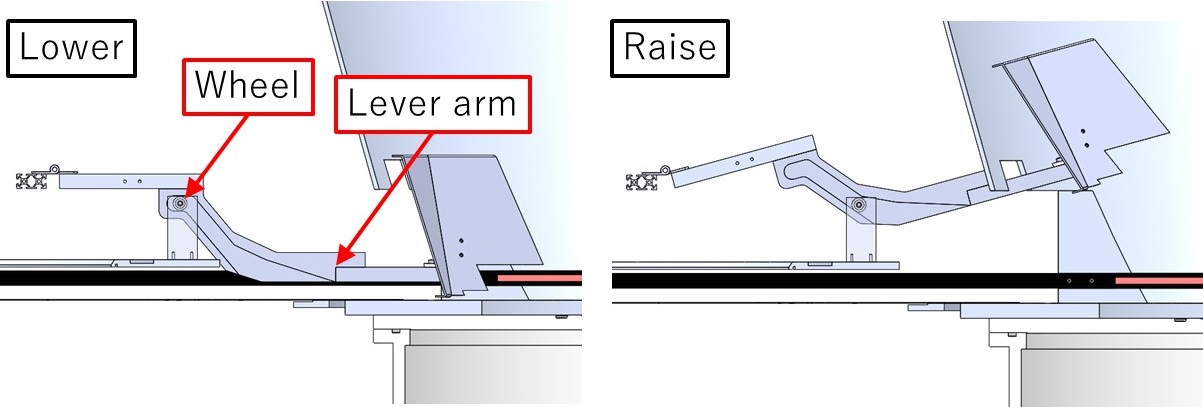}
        \subcaption{}
        \label{fig:shutter_groove}
    \end{minipage}
    \\
     \begin{minipage}{0.48\textwidth}%
        \centering
        \includegraphics[height=3.5cm]{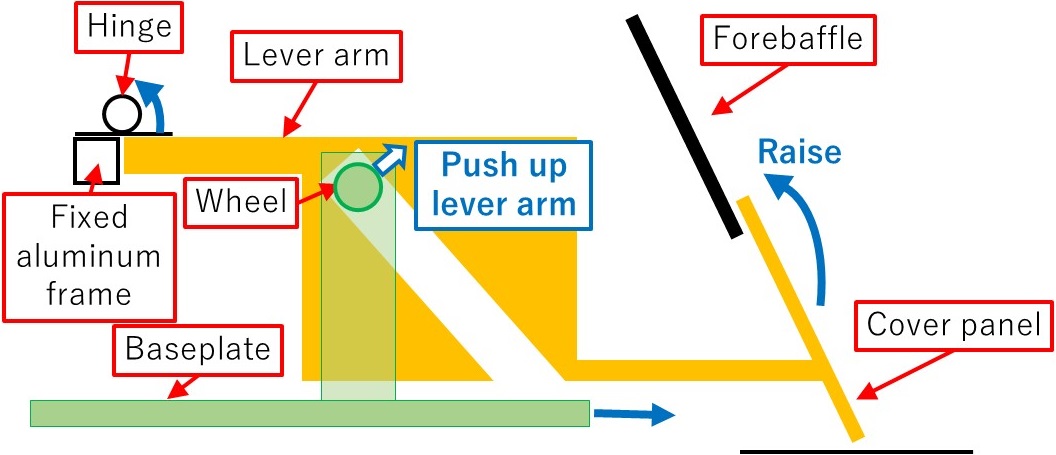}
        \subcaption{}
        \label{fig:shutter_cross_section}
    \end{minipage}
    \caption{
    (\subref{fig:shutter_overview}) Forebaffle opening with the cover panel shown from the inside of the forebaffle.
    (\subref{fig:shutter_groove}) side-views of the lever arm when the cover panel is lowered and raised, and
    (\subref{fig:shutter_cross_section}) components of the cover system in the sideview. When inserting and removing the SWG assembly, the wheels fixed on the baseplate push the lever arms up and down to open and close the cover panel.
    }
    \label{fig:shutter}
\end{figure}

The SWG is moved by the actuators through the forebaffle opening. The opening has a potential risk of making optical anisotropy for the detectors.
Therefore, we developed a system to cover the opening during the CMB observation. 
The cover system consists of a cover panel that fits into the opening shape and two lever arms to raise and lower the panel as shown in Fig.~\ref{fig:shutter}.
The cover panel and the inner surface of the forebaffle are blackened with a microwave absorber, Laird HR10~\cite{Laird}, to prevent ambient light from reflecting on the surface.
Therefore, the surface of the cover panel is optically the same as the surface of the forebaffle.
The panel rises and falls in synchronizing with the motions of inserting and removing the SWG assembly into the forebaffle by the actuators, respectively.
The SWG assembly has wheels fixed on the baseplate as shown in Fig.~\ref{fig:wire_grid_assembly} and the lever arms have grooves, in which the wheels run.
When inserting and removing the SWG assembly, the wheels push the lever arms up and down to open and close the cover panel.
To open and close the cover, 5 kg force is required on the actuators.
This force is included in the $23\,\mathrm{kg}$ of requirement on the actuator loadable weight.

\subsection{Tests for the actuators and operation}
We evaluated the loadable weight of the actuators by putting a dummy mass on the actuators tilted vertically. 
We concluded that the two actuators can carry up to $\sim 50\,\mathrm{kg}$-load vertically.
We also measured the operation time.
To obtain calibration data at each wire angle, 
a time interval of 10 seconds was set between every $22.5^{\circ}$ rotations.~\cite{Kusaka2014}
We found that each rotation step was positioned with an accuracy of $0.6^{\circ}$.
The insertion and removal time was 1.5 minutes and the total operation time was 7.5 minutes.
This operation time will allow us to perform the angle calibration a few times per day or more.

\section{Evaluation of the angle systematic error}\label{sec:meas}
We evaluated whether the developed calibration system satisfies the requirement of the angle accuracy to achieve the SO science target.
The uncertainties in the wire angle come from the alignment accuracy of the wires on the ring and measurement systematic errors with the encoder and gravity sensor.
The wire alignment is defined by the wire groove as described in Section~\ref{sec:design}.
The wire alignment is determined by the groove width of $0.2\,\mathrm{mm}$, which corresponds to angle uncertainty of $0.02^{\circ}$, 
and the machining accuracy of the groove is $0.05\,\mathrm{mm}$, which corresponds to an angle uncertainty of $0.005^{\circ}$.

\begin{figure}
    \begin{minipage}{0.48\textwidth}%
        \centering
        \includegraphics[width=7cm]{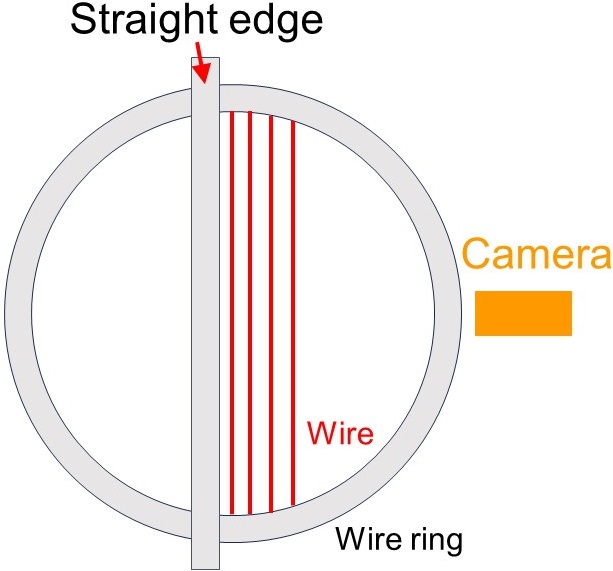}%
        \subcaption{}
        \label{fig:sag_setup_a}
    \end{minipage}
    \\
    \begin{minipage}{0.48\textwidth}
        \centering
        \includegraphics[width=7cm]{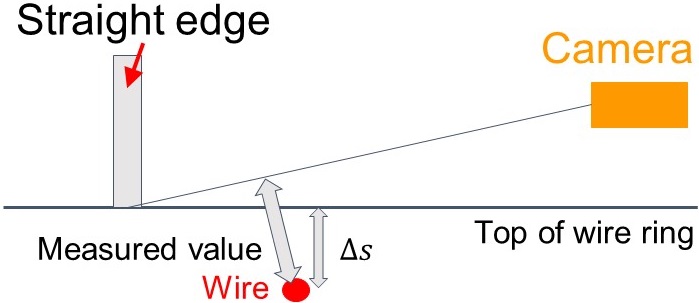}%
        \subcaption{}
        \label{fig:sag_setup_b}
    \end{minipage}
    \caption{
    (\subref{fig:sag_setup_a}) Top view of the setup for wire sag measurement, (\subref{fig:sag_setup_b}) cross section of the setup. 
    $\Delta s$ is the sag length and is converted from the measured value using the positions of the camera and the straight edge.
    }
    \label{fig:sag_setup}
\end{figure}

Wire sagging is another possible cause of wire misalignment.
We took a photo of each wire with a commercial camera from the side of the wire when the SWG was put on a horizontal plane to measure the length of sagging from the horizontal at the center of the wire (Fig.~\ref{fig:sag_setup}). As the horizontal plane, a precision steel straight edge (Ohnishi OS-140-1000A~\cite{Ohnishi}) was placed on the SWG ring. 
The sag length $\Delta s$ was converted from the measured value using the positions of the camera and the straight edge. The measured value was calculated by counting the number of pixels in the photo between the wire and the edge of the straight edge. To convert the number of pixels to millimeters, a ruler was also placed in the photo.


The average $\Delta s$ for all 39 wires was $0.10\pm0.14\,\mathrm{mm}$. The measurement error of $\Delta s$ comes from four factors: 
(a) the straightness of the straight edge, (b) the pixel size of the camera, (c) the uncertainties of the positions of the camera, and (d) the positions of the straight edge.
Factor (a) comes from the machining accuracy and corresponds to $0.03\,\mathrm{mm}$ in $\Delta s$. The (b) corresponds to $0.05\,\mathrm{mm}$. We took half of the pixel size as a measurement error, which corresponds to $0.025\,\mathrm{mm}$. Concerning factor (c), we conservatively assumed that the position accuracy of the camera focus is the radius of the lens, which corresponds to $0.05\,\mathrm{mm}$ in $\Delta s$. The (d) is measured with a ruler, and it is conservatively assumed that the position accuracy of the straight edge is a length of one scale of this ruler. This accuracy corresponds to $0.06\,\mathrm{mm}$ in $\Delta s$. We concluded that the total error for measuring $\Delta s$ is $0.1\,\mathrm{mm}$.
The sag angle was calculated by $\Delta\theta = \arctan\left(\Delta s/r\right)$, where $r$ is half the length of each wire.
$\Delta\theta$ corresponds to the sag angle when the SWG is placed vertically, which is the maximum angle variation caused by sagging.
The average sag angle in all 39 wires was $0.02^{\circ}\pm0.03^{\circ}$.
Thus, the angle misalignment due to the sag was $<0.05^{\circ}$.

The systematic error of the rotation angle measured by the encoder was determined by not only the resolution but also the alignment of the magnetic scale attached to the SWG bearing.
We measured the actual rotation angle using a coordinate measuring machine (Faro Edge~\cite{Faro}). 
The accuracy of the coordinate measuring machine is $29\ \mu \mbox{m}$, which corresponds to the angle systematic error of $0.003^{\circ}$.
With the coordinate measuring machine, we measured the longitudinal direction of the two special grooves dug on the SWG parallel to the direction of the wires instead of measuring the wires directly. The alignment between the grooves and wires was confirmed by the machining accuracy, which is equal to $0.02^{\circ}$ in the rotation angle.
We did 12 rotations with random angles and measured the angle of wires with the coordinate measuring machine. At the same time, the encoder measured the rotation angle as well. We found that the fluctuation of the angle differences between the coordinate measuring machine and encoder was independent of wire angles and that the standard deviation was $0.03^{\circ}$ in the 12 measurements. Therefore, we concluded that this systematic error was $0.03^{\circ}$ in the angle calibration.
As shown in Eq.~\ref{eq:angle_calc}, we have to measure $\theta_{\mathrm{enc}}, \theta_{\mathrm{enc}0}$, and $\theta_{\mathrm{sens}} = \mathrm{arctan}(\mathrm{sin} \alpha/\mathrm{sin} \beta)$ to monitor $\theta_{\mathrm{wire}}$. $\theta_{\mathrm{sens}}$ can be monitored with a systematic error of $0.02^{\circ}$ 
because the measurement accuracy of the gravity sensor to measure $\alpha$ and $\beta$ are $0.01^{\circ}$ and $\alpha$ and $\beta$ were confirmed to be around $0^{\circ}$ and $40^{\circ}$ during the calibration respectively.
$\theta_{\mathrm{enc}0}$ was evaluated by measuring $\theta_{\mathrm{enc}}, \alpha$, and  $\beta$ at $\theta_{\mathrm{wire}} = 0^{\circ}$, which was confirmed by an extra precision level meter (SK FLW-200002) with an accuracy of $0.002^{\circ}$.
We found that the uncertainty of $\theta_{\mathrm{enc}0}$ was $0.04^{\circ}$. The uncertainty consists of the encoder systematic error of $0.03^{\circ}$ as evaluated in the coordinate measuring machine test, the $\theta_{\mathrm{sens}}$ systematic error of $0.02^\circ$, and the machining accuracy of the wire grooves, which corresponds to $0.002^{\circ}$ in $\theta_{\mathrm{wire}}$.

\begin{table}[h!]
\renewcommand{\footnoterule}{\empty} 
\renewcommand{\thefootnote}{\alph{footnote}} 
\begin{minipage}{0.48\textwidth} 
\begin{center}
\caption{Expected systematic error in the global angle calibration. The total expected systematic error is less than the target angle accuracy of $0.1^{\circ}$.}
\label{tab:achievement}
\begin{tabular}{|c|c|} \hline
Source of inaccuracy & Systematic error \\ \hline
Wire alignment & $0.02^{\circ}$ \\
Wire sag & $<0.05^{\circ}$\\
Encoder & $0.03^{\circ}$\\
Encoder reference & $<0.04^{\circ}$\\
Gravity sensor & $<0.02^{\circ}$\\\hline
Total & $0.08^{\circ}$\footnotemark[1]\\ \hline 
\end{tabular}
\footnotetext[1]{The total systematic error is obtained by quadrature sum of each systematic error. We assume that these uncertainties are uncorrelated.}
\end{center}
\end{minipage} 
\end{table}

Table~\ref{tab:achievement} summarizes the evaluation results. We concluded that the total expected systematic error is $0.08^{\circ}$.

\section{Conclusion}
We report the development of an angle calibration system with a sparse wire grid (SWG) for SAT receivers in the SO experiment.
One of the new features of the calibration system is a fully remote-controlled system for regular calibration, which is enabled by the two electric actuators switching the SWG position between the calibration time and the CMB observation. 
This fully remote-controlled system allows us to perform the angle calibration a few times per day or more instead of several times per season in the past experiments.
In addition, the calibration system has a sufficiently small expected systematic error on the wire angle, which is achieved by the system monitoring the wire orientation with an encoder and gravity sensor. 
We have designed the calibration system to satisfy the size requirements to integrate the system with the SAT receiver.

The angle systematic error and the loadable weight of the traveling system have been evaluated with the developed calibration system.
In particular, the expected systematic error in the global angle calibration is $0.08^{\circ}$.

\begin{acknowledgments}
This work was supported by the Simons Foundation (Award $\#$457687, B.K.). This work was supported in part by World Premier International Research Center Initiative (WPI Initiative), MEXT, Japan. In Japan, this work was supported by JSPS KAKENHI Grant Numbers JP17H06134, JP18J01039, JP19H00674, JP22H04913, JP23H00105, and the JSPS Core-to-Core Program JPJSCCA20200003. This work was also supported by JST, the establishment of university fellowships towards the creation of science technology innovation, Grant Number JPMJFS2108. Work at LBNL is supported in part by the U.S. Department of Energy, Office of Science,Office of High Energy Physics, under contract No. DE-AC02-05CH11231. MM is supported by IIW fellowship. FM acknowledges JSPS KAKENHI Grant Number 20K14483.
\end{acknowledgments}

\tocless{\section*{Author Declarations}}

\tocless{}{\subsection*{Conflict of Interest}}
The authors have no conflicts to disclose.\\
\tocless{}{\subsection*{Author Contributions}}
\textbf{Masaaki Murata}: Conceptualization (equal); Investigation (equal); Methodology (equal); Validation (equal); Visualization (equal); Writing – original draft (equal);.
\textbf{Hironobu Nakata}: Data Curation (equal); Investigation (equal); Methodology (equal); Project Administration (equal); Software (equal); Visualization (equal); Writing – original draft (equal).
\textbf{Kengo Iijima}: Data Curation (equal); Formal Analysis (equal); Investigation (equal); Methodology (equal); Software (equal); Visualization (equal); Writing – original draft (equal).
\textbf{Shunsuke Adachi}: Conceptualization (equal); Investigation (equal); Methodology (equal); Resources (equal); Software (equal); Writing – original draft (equal).
\textbf{Yudai Seino}: Investigation (equal).
\textbf{Kenji Kiuchi}: Methodology (equal).
\textbf{Frederick Matsuda}: Methodology (equal).
\textbf{Michael J. Randall}: Investigation (equal); Resources (equal); Software (equal); Validation (equal).
\textbf{Kam Arnold}: Funding Acquisition (supporting); Supervision (supporting).
\textbf{Nicholas Galitzki}: Conceptualization (supporting); Supervision (supporting); Writing – review $\&$ editing (supporting).
\textbf{Bradley R. Johnson}: Funding Acquisition (supporting); Supervision (supporting).
\textbf{Brian Keating}: Funding Acquisition (supporting); Supervision (supporting).
\textbf{Akito Kusaka}: Conceptualization (supporting); Funding Acquisition (supporting); Methodology (supporting); Project Administration (supporting).
\textbf{John B. Lloyd}: Investigation (supporting).
\textbf{Joseph Seibert}: Validation (supporting).
\textbf{Maximiliano Silva-Feaver}: Investigation (supporting); Resources (supporting); Writing – review $\&$ editing (supporting).
\textbf{Osamu Tajima}: Conceptualization (supporting); Funding Acquisition (supporting); Methodology (supporting); Project Administration (supporting).
\textbf{Tomoki Terasaki}: Validation (supporting); Writing – review $\&$ editing (supporting).
\textbf{Kyohei Yamada}: Formal Analysis (supporting); Writing – review $\&$ editing (supporting).
\\
\tocless{\section*{Data availability}}
The data that support the findings of this study are available from the corresponding author upon reasonable request.

\appendix
\section{Cross polarization}
\label{sec:cross_pol}

\begin{figure}
\centering\includegraphics[width=8cm]{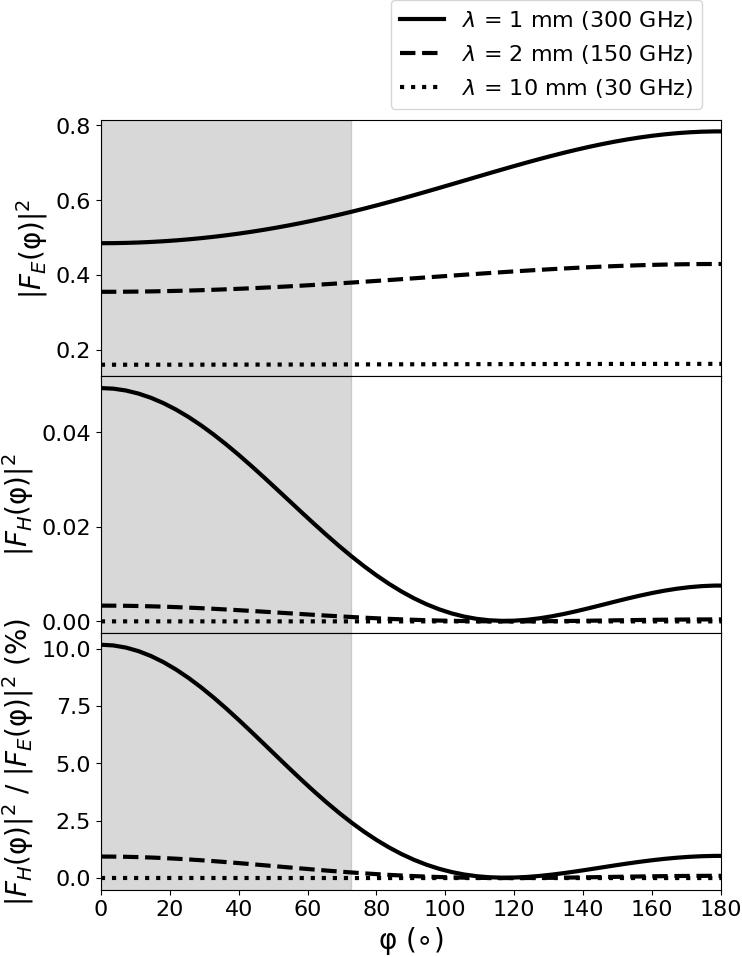}
\caption{Scattering coefficient and ratio of the scattering power of H-polarization to that of E-polarization as a function of scattering angle $\varphi$. $\lambda$ is varied over the range of the observation wavelength of the SATs and the wire radius $a$ is $0.05\,\mathrm{mm}$. The grey area means $\varphi < 72.5^{\circ}$.} 
\label{fig:ratio_E_H}
\end{figure}

Here we derive the ratio of the power of the cross-polarization signal to that of the polarization signal parallel to the wires of the SWG. We consider a simple model with a perfectly conducting wire with radius $a$ along the z-axis and light is incident along the x-axis and scattered on the surface of the wire. 
We assume the incident light has one of two polarizations, with the electric field in either the z-axis direction (i.e. parallel to the wire), which we call ``E-polarization'', or the y-axis direction (i.e. perpendicular to the wire), which we call ``H-polarization''. E-polarization and H-polarization generate polarization signal parallel to the wire and cross-polarization signal relative to the wire direction respectively.
The scattered field in the far zone are expressed as~\cite{Osipov2017}

\begin{equation}
u (r, \varphi) = - \sqrt{\frac{2}{\pi kr}}\mathrm{e}^{-j(kr-\frac{\pi}{4})}F(\varphi),
\end{equation}
where $k = 2\pi / \lambda$, $\lambda$ is the wavelength of the incident light, $r$ is a distance from the center of the wire,
and $\varphi$ is the angle between the propagation directions of the scattered light and the incident light around the wire. When the scattered light propagates in the direction from which the incident light came, $\varphi = 0^{\circ}$. On the other hand, when the scattered light propagates in the direction to which the incident light goes, $\varphi = 180^{\circ}$.
$F(\varphi)$ is called as ``scattering coefficient'' of the wire. The scattering coefficients for E-polarization and H-polarization are expressed as
\begin{align}
    F_{\mathrm{E}}(\varphi) &= \sum^{\infty}_{n=-\infty} \frac{J_n(ka)}{H^{(2)}_n(ka)}\mathrm{e}^{-jn(\varphi - \pi)}\\
    F_{\mathrm{H}}(\varphi) &= \sum^{\infty}_{n=-\infty} \frac{J^{\prime}_n(ka)}{H^{(2)\prime}_n(ka)}\mathrm{e}^{-jn(\varphi - \pi)},
\end{align}
where $J_n(ka)$ is the Bessel function of the first kind and $H^{(2)}_n(ka)$ is the Hankel function of the second kind. $J^{\prime}_n(ka)$ and $H^{(2)\prime}_n(ka)$ are the derivatives of $J_n(ka)$ and $H^{(2)}_n(ka)$ with respect to $ka$ respectively. The scattering power of E-polarization and H-polarization are $|u_{\mathrm{E}}(r, \varphi)|^2$ and $|u_{\mathrm{H}}(r, \varphi)|^2$, respectively, and their ratio, $|u_{\mathrm{H}}(r, \varphi)|^2 / |u_{\mathrm{E}}(r, \varphi)|^2$, equals to $|F_{\mathrm{H}}(\varphi)|^2 / |F_{\mathrm{E}}(\varphi)|^2$.
Figure~\ref{fig:ratio_E_H} shows the scattering coefficients and the ratios of the scattering power of E-polarization to that of H-polarization with $\lambda$ varying over the range of the observation wavelength of the SATs, and $a=0.05\,\mathrm{mm}$. 
We find that the ratio scales with wavelength as $\sim\lambda^{-3.8}$.

\begin{figure}
\centering\includegraphics[height=5cm]{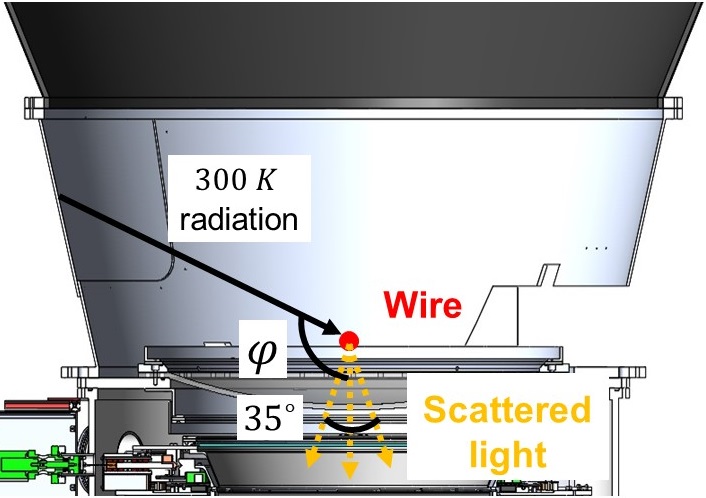}
\caption{A single wire of the SWG in the SAT. The longitudinal direction of the wire is perpendicular to the paper. The scattered light entering the detectors located below this figure has a $\varphi$ larger than $90^{\circ} - 17.5^{\circ} = 72.5^{\circ}$ because the field-of-view of the SAT is $35^{\circ}$}
\label{fig:SWG_E_H}
\end{figure}

Next, we consider the situation of the SWG in a SAT as shown in Fig.~\ref{fig:SAT_cross_section} and Fig.~\ref{fig:-20dBcone}. Since the wire pitch of the SWG is much larger than the wavelength of the observed light, the fraction of cross polarization can be well approximated by that of a single wire.
The $300\mathrm{\,K}$ blackbody radiation from the forebaffle is incident and scattered on the surface of the wire. Considering the field-of-view of the SAT and the position of the SWG and the forebaffle as shown in Fig.~\ref{fig:SWG_E_H}, the scattered light entering the detectors has a $\varphi$ larger than $72.5^{\circ}$. When integrated over this range, the total cross-polarization signal is smaller than $\sim 1\%$ of the total polarization signal even for 300 GHz, which is the worst case here.

\bibliography{Main}

\begin{thebibliography}{53}%
\makeatletter
\providecommand \@ifxundefined [1]{%
 \@ifx{#1\undefined}
}%
\providecommand \@ifnum [1]{%
 \ifnum #1\expandafter \@firstoftwo
 \else \expandafter \@secondoftwo
 \fi
}%
\providecommand \@ifx [1]{%
 \ifx #1\expandafter \@firstoftwo
 \else \expandafter \@secondoftwo
 \fi
}%
\providecommand \natexlab [1]{#1}%
\providecommand \enquote  [1]{``#1''}%
\providecommand \bibnamefont  [1]{#1}%
\providecommand \bibfnamefont [1]{#1}%
\providecommand \citenamefont [1]{#1}%
\providecommand \href@noop [0]{\@secondoftwo}%
\providecommand \href [0]{\begingroup \@sanitize@url \@href}%
\providecommand \@href[1]{\@@startlink{#1}\@@href}%
\providecommand \@@href[1]{\endgroup#1\@@endlink}%
\providecommand \@sanitize@url [0]{\catcode `\\12\catcode `\$12\catcode
  `\&12\catcode `\#12\catcode `\^12\catcode `\_12\catcode `\%12\relax}%
\providecommand \@@startlink[1]{}%
\providecommand \@@endlink[0]{}%
\providecommand \url  [0]{\begingroup\@sanitize@url \@url }%
\providecommand \@url [1]{\endgroup\@href {#1}{\urlprefix }}%
\providecommand \urlprefix  [0]{URL }%
\providecommand \Eprint [0]{\href }%
\providecommand \doibase [0]{http://dx.doi.org/}%
\providecommand \selectlanguage [0]{\@gobble}%
\providecommand \bibinfo  [0]{\@secondoftwo}%
\providecommand \bibfield  [0]{\@secondoftwo}%
\providecommand \translation [1]{[#1]}%
\providecommand \BibitemOpen [0]{}%
\providecommand \bibitemStop [0]{}%
\providecommand \bibitemNoStop [0]{.\EOS\space}%
\providecommand \EOS [0]{\spacefactor3000\relax}%
\providecommand \BibitemShut  [1]{\csname bibitem#1\endcsname}%
\let\auto@bib@innerbib\@empty
\bibitem [{\citenamefont {{Planck
  Collaboration}}(2020{\natexlab{a}})}]{Planck_cosmo_2020}%
  \BibitemOpen
  \bibfield  {author} {\bibinfo {author} {\bibnamefont {{Planck
  Collaboration}}},\ }\bibfield  {title} {\enquote {\bibinfo {title} {{Planck
  2018 results - VI. Cosmological parameters}},}\ }\href {\doibase
  10.1051/0004-6361/201833910} {\bibfield  {journal} {\bibinfo  {journal}
  {A\&A}\ }\textbf {\bibinfo {volume} {641}},\ \bibinfo {pages} {A6} (\bibinfo
  {year} {2020}{\natexlab{a}})}\BibitemShut {NoStop}%
\bibitem [{\citenamefont {{Planck
  Collaboration}}(2020{\natexlab{b}})}]{Planck_infla_2020}%
  \BibitemOpen
  \bibfield  {author} {\bibinfo {author} {\bibnamefont {{Planck
  Collaboration}}},\ }\bibfield  {title} {\enquote {\bibinfo {title} {{Planck
  2018 results - X. Constraints on inflation}},}\ }\href {\doibase
  10.1051/0004-6361/201833887} {\bibfield  {journal} {\bibinfo  {journal}
  {A\&A}\ }\textbf {\bibinfo {volume} {641}},\ \bibinfo {pages} {A10} (\bibinfo
  {year} {2020}{\natexlab{b}})}\BibitemShut {NoStop}%
\bibitem [{\citenamefont {Guth}(1981)}]{Guth1981}%
  \BibitemOpen
  \bibfield  {author} {\bibinfo {author} {\bibfnamefont {A.~H.}\ \bibnamefont
  {Guth}},\ }\bibfield  {title} {\enquote {\bibinfo {title} {{Inflationary
  universe: A possible solution to the horizon and flatness problems}},}\
  }\href {\doibase 10.1103/PhysRevD.23.347} {\bibfield  {journal} {\bibinfo
  {journal} {Phys. Rev. D}\ }\textbf {\bibinfo {volume} {23}},\ \bibinfo
  {pages} {347--356} (\bibinfo {year} {1981})}\BibitemShut {NoStop}%
\bibitem [{\citenamefont {Carroll}(1998{\natexlab{a}})}]{Seljak1997}%
  \BibitemOpen
  \bibfield  {author} {\bibinfo {author} {\bibfnamefont {S.~M.}\ \bibnamefont
  {Carroll}},\ }\bibfield  {title} {\enquote {\bibinfo {title} {{Quintessence
  and the Rest of the World: Suppressing Long-Range Interactions}},}\ }\href
  {\doibase 10.1103/PhysRevLett.81.3067} {\bibfield  {journal} {\bibinfo
  {journal} {Phys. Rev. Lett.}\ }\textbf {\bibinfo {volume} {81}},\ \bibinfo
  {pages} {3067--3070} (\bibinfo {year} {1998}{\natexlab{a}})}\BibitemShut
  {NoStop}%
\bibitem [{\citenamefont {Keating}, \citenamefont {Shimon},\ and\ \citenamefont
  {Yadav}(2012)}]{Keating2013}%
  \BibitemOpen
  \bibfield  {author} {\bibinfo {author} {\bibfnamefont {B.~G.}\ \bibnamefont
  {Keating}}, \bibinfo {author} {\bibfnamefont {M.}~\bibnamefont {Shimon}}, \
  and\ \bibinfo {author} {\bibfnamefont {A.~P.~S.}\ \bibnamefont {Yadav}},\
  }\bibfield  {title} {\enquote {\bibinfo {title} {{SELF-CALIBRATION OF COSMIC
  MICROWAVE BACKGROUND POLARIZATION EXPERIMENTS}},}\ }\href {\doibase
  10.1088/2041-8205/762/2/L23} {\bibfield  {journal} {\bibinfo  {journal} {The
  Astrophysical Journal Letters}\ }\textbf {\bibinfo {volume} {762}},\ \bibinfo
  {pages} {L23} (\bibinfo {year} {2012})}\BibitemShut {NoStop}%
\bibitem [{\citenamefont {Kaufman}\ \emph {et~al.}(2014)\citenamefont
  {Kaufman}, \citenamefont {Miller}, \citenamefont {Shimon} \emph
  {et~al.}}]{Kaufman2014}%
  \BibitemOpen
  \bibfield  {author} {\bibinfo {author} {\bibfnamefont {J.~P.}\ \bibnamefont
  {Kaufman}}, \bibinfo {author} {\bibfnamefont {N.~J.}\ \bibnamefont {Miller}},
  \bibinfo {author} {\bibfnamefont {M.}~\bibnamefont {Shimon}},  \emph
  {et~al.},\ }\bibfield  {title} {\enquote {\bibinfo {title} {{Self-calibration
  of BICEP1 three-year data and constraints on astrophysical polarization
  rotation}},}\ }\href {\doibase 10.1103/PhysRevD.89.062006} {\bibfield
  {journal} {\bibinfo  {journal} {Phys. Rev. D}\ }\textbf {\bibinfo {volume}
  {89}},\ \bibinfo {pages} {062006} (\bibinfo {year} {2014})}\BibitemShut
  {NoStop}%
\bibitem [{\citenamefont {Tristram}\ \emph {et~al.}(2022)\citenamefont
  {Tristram}, \citenamefont {Banday}, \citenamefont {G\'orski} \emph
  {et~al.}}]{Tristram2022}%
  \BibitemOpen
  \bibfield  {author} {\bibinfo {author} {\bibfnamefont {M.}~\bibnamefont
  {Tristram}}, \bibinfo {author} {\bibfnamefont {A.~J.}\ \bibnamefont
  {Banday}}, \bibinfo {author} {\bibfnamefont {K.~M.}\ \bibnamefont
  {G\'orski}},  \emph {et~al.},\ }\bibfield  {title} {\enquote {\bibinfo
  {title} {{Improved limits on the tensor-to-scalar ratio using BICEP and
  $Planck$ data}},}\ }\href {\doibase 10.1103/PhysRevD.105.083524} {\bibfield
  {journal} {\bibinfo  {journal} {Phys. Rev. D}\ }\textbf {\bibinfo {volume}
  {105}},\ \bibinfo {pages} {083524} (\bibinfo {year} {2022})}\BibitemShut
  {NoStop}%
\bibitem [{\citenamefont {Carroll}(1998{\natexlab{b}})}]{Carroll1998}%
  \BibitemOpen
  \bibfield  {author} {\bibinfo {author} {\bibfnamefont {S.~M.}\ \bibnamefont
  {Carroll}},\ }\bibfield  {title} {\enquote {\bibinfo {title} {{Quintessence
  and the Rest of the World: Suppressing Long-Range Interactions}},}\ }\href
  {\doibase 10.1103/PhysRevLett.81.3067} {\bibfield  {journal} {\bibinfo
  {journal} {Phys. Rev. Lett.}\ }\textbf {\bibinfo {volume} {81}},\ \bibinfo
  {pages} {3067--3070} (\bibinfo {year} {1998}{\natexlab{b}})}\BibitemShut
  {NoStop}%
\bibitem [{\citenamefont {Jost}, \citenamefont {Errard},\ and\ \citenamefont
  {Stompor}(2023)}]{Jost2023}%
  \BibitemOpen
  \bibfield  {author} {\bibinfo {author} {\bibfnamefont {B.}~\bibnamefont
  {Jost}}, \bibinfo {author} {\bibfnamefont {J.}~\bibnamefont {Errard}}, \ and\
  \bibinfo {author} {\bibfnamefont {R.}~\bibnamefont {Stompor}},\ }\href@noop
  {} {\enquote {\bibinfo {title} {{Characterising cosmic birefringence in the
  presence of galactic foregrounds and instrumental systematic effects}},}\ }
  (\bibinfo {year} {2023}),\ \Eprint {http://arxiv.org/abs/2212.08007}
  {arXiv:2212.08007 [astro-ph.CO]} \BibitemShut {NoStop}%
\bibitem [{\citenamefont {{The Simons Observatory
  collaboration}}(2019)}]{Ade2019}%
  \BibitemOpen
  \bibfield  {author} {\bibinfo {author} {\bibnamefont {{The Simons Observatory
  collaboration}}},\ }\bibfield  {title} {\enquote {\bibinfo {title} {{The
  Simons Observatory: science goals and forecasts}},}\ }\href {\doibase
  10.1088/1475-7516/2019/02/056} {\bibfield  {journal} {\bibinfo  {journal}
  {Journal of Cosmology and Astroparticle Physics}\ }\textbf {\bibinfo {volume}
  {2019}},\ \bibinfo {pages} {056} (\bibinfo {year} {2019})}\BibitemShut
  {NoStop}%
\bibitem [{\citenamefont {Bryan}\ \emph {et~al.}(2018)\citenamefont {Bryan},
  \citenamefont {Simon}, \citenamefont {Gerbino} \emph {et~al.}}]{Bryan2018}%
  \BibitemOpen
  \bibfield  {author} {\bibinfo {author} {\bibfnamefont {S.~A.}\ \bibnamefont
  {Bryan}}, \bibinfo {author} {\bibfnamefont {S.~M.}\ \bibnamefont {Simon}},
  \bibinfo {author} {\bibfnamefont {M.}~\bibnamefont {Gerbino}},  \emph
  {et~al.},\ }\bibfield  {title} {\enquote {\bibinfo {title} {{Development of
  calibration strategies for the Simons Observatory}},}\ }in\ \href {\doibase
  10.1117/12.2313832} {\emph {\bibinfo {booktitle} {Millimeter, Submillimeter,
  and Far-Infrared Detectors and Instrumentation for Astronomy IX}}},\ Vol.\
  \bibinfo {volume} {10708},\ \bibinfo {editor} {edited by\ \bibinfo {editor}
  {\bibfnamefont {J.}~\bibnamefont {Zmuidzinas}}\ and\ \bibinfo {editor}
  {\bibfnamefont {J.-R.}\ \bibnamefont {Gao}}},\ \bibinfo {organization}
  {International Society for Optics and Photonics}\ (\bibinfo  {publisher}
  {SPIE},\ \bibinfo {year} {2018})\ p.\ \bibinfo {pages} {1070840}\BibitemShut
  {NoStop}%
\bibitem [{\citenamefont {Abitbol}\ \emph {et~al.}(2021)\citenamefont
  {Abitbol}, \citenamefont {Alonso}, \citenamefont {Simon} \emph
  {et~al.}}]{Abitbol2021}%
  \BibitemOpen
  \bibfield  {author} {\bibinfo {author} {\bibfnamefont {M.~H.}\ \bibnamefont
  {Abitbol}}, \bibinfo {author} {\bibfnamefont {D.}~\bibnamefont {Alonso}},
  \bibinfo {author} {\bibfnamefont {S.~M.}\ \bibnamefont {Simon}},  \emph
  {et~al.},\ }\bibfield  {title} {\enquote {\bibinfo {title} {{The Simons
  Observatory: gain, bandpass and polarization-angle calibration requirements
  for B-mode searches}},}\ }\href {\doibase 10.1088/1475-7516/2021/05/032}
  {\bibfield  {journal} {\bibinfo  {journal} {Journal of Cosmology and
  Astroparticle Physics}\ }\textbf {\bibinfo {volume} {2021}},\ \bibinfo
  {pages} {032} (\bibinfo {year} {2021})}\BibitemShut {NoStop}%
\bibitem [{\citenamefont {Johnson}\ \emph {et~al.}(2015)\citenamefont
  {Johnson}, \citenamefont {Vourch}, \citenamefont {Drysdale} \emph
  {et~al.}}]{Johnson2015}%
  \BibitemOpen
  \bibfield  {author} {\bibinfo {author} {\bibfnamefont {B.~R.}\ \bibnamefont
  {Johnson}}, \bibinfo {author} {\bibfnamefont {C.~J.}\ \bibnamefont {Vourch}},
  \bibinfo {author} {\bibfnamefont {T.~D.}\ \bibnamefont {Drysdale}},  \emph
  {et~al.},\ }\bibfield  {title} {\enquote {\bibinfo {title} {{A CubeSat for
  Calibrating Ground-Based and Sub-Orbital Millimeter-Wave Polarimeters
  (CalSat)}},}\ }\href {\doibase 10.1142/S2251171715500075} {\bibfield
  {journal} {\bibinfo  {journal} {Journal of Astronomical Instrumentation}\
  }\textbf {\bibinfo {volume} {04}},\ \bibinfo {pages} {1550007} (\bibinfo
  {year} {2015})}\BibitemShut {NoStop}%
\bibitem [{\citenamefont {Takahashi}\ \emph {et~al.}(2008)\citenamefont
  {Takahashi}, \citenamefont {Barkats}, \citenamefont {Battle} \emph
  {et~al.}}]{Takahashi2008}%
  \BibitemOpen
  \bibfield  {author} {\bibinfo {author} {\bibfnamefont {Y.~D.}\ \bibnamefont
  {Takahashi}}, \bibinfo {author} {\bibfnamefont {D.}~\bibnamefont {Barkats}},
  \bibinfo {author} {\bibfnamefont {J.~O.}\ \bibnamefont {Battle}},  \emph
  {et~al.},\ }\bibfield  {title} {\enquote {\bibinfo {title} {{CMB polarimetry
  with BICEP: instrument characterization, calibration, and performance}},}\
  }in\ \href {\doibase 10.1117/12.790306} {\emph {\bibinfo {booktitle}
  {Millimeter and Submillimeter Detectors and Instrumentation for Astronomy
  IV}}},\ Vol.\ \bibinfo {volume} {7020},\ \bibinfo {editor} {edited by\
  \bibinfo {editor} {\bibfnamefont {W.~D.}\ \bibnamefont {Duncan}}, \bibinfo
  {editor} {\bibfnamefont {W.~S.}\ \bibnamefont {Holland}}, \bibinfo {editor}
  {\bibfnamefont {S.}~\bibnamefont {Withington}}, \ and\ \bibinfo {editor}
  {\bibfnamefont {J.}~\bibnamefont {Zmuidzinas}}},\ \bibinfo {organization}
  {International Society for Optics and Photonics}\ (\bibinfo  {publisher}
  {SPIE},\ \bibinfo {year} {2008})\ p.\ \bibinfo {pages} {70201D}\BibitemShut
  {NoStop}%
\bibitem [{\citenamefont {{The Polarbear
  Collaboration}}(2014)}]{Polarbear2014}%
  \BibitemOpen
  \bibfield  {author} {\bibinfo {author} {\bibnamefont {{The Polarbear
  Collaboration}}},\ }\bibfield  {title} {\enquote {\bibinfo {title} {{A
  MEASUREMENT OF THE COSMIC MICROWAVE BACKGROUND B-MODE POLARIZATION POWER
  SPECTRUM AT SUB-DEGREE SCALES WITH POLARBEAR}},}\ }\href {\doibase
  10.1088/0004-637X/794/2/171} {\bibfield  {journal} {\bibinfo  {journal} {The
  Astrophysical Journal}\ }\textbf {\bibinfo {volume} {794}},\ \bibinfo {pages}
  {171} (\bibinfo {year} {2014})}\BibitemShut {NoStop}%
\bibitem [{\citenamefont {{QUIET Collaboration}}(2011)}]{Bischoff2011}%
  \BibitemOpen
  \bibfield  {author} {\bibinfo {author} {\bibnamefont {{QUIET
  Collaboration}}},\ }\bibfield  {title} {\enquote {\bibinfo {title} {{FIRST
  SEASON QUIET OBSERVATIONS: MEASUREMENTS OF COSMIC MICROWAVE BACKGROUND
  POLARIZATION POWER SPECTRA AT 43 GHz IN THE MULTIPOLE RANGE 25 < $\ell$ <
  475}},}\ }\href {\doibase 10.1088/0004-637X/741/2/111} {\bibfield  {journal}
  {\bibinfo  {journal} {The Astrophysical Journal}\ }\textbf {\bibinfo {volume}
  {741}},\ \bibinfo {pages} {111} (\bibinfo {year} {2011})}\BibitemShut
  {NoStop}%
\bibitem [{\citenamefont {O'Dell}, \citenamefont {Swetz},\ and\ \citenamefont
  {Timbie}(2002)}]{ODell2002}%
  \BibitemOpen
  \bibfield  {author} {\bibinfo {author} {\bibfnamefont {C.}~\bibnamefont
  {O'Dell}}, \bibinfo {author} {\bibfnamefont {D.}~\bibnamefont {Swetz}}, \
  and\ \bibinfo {author} {\bibfnamefont {P.}~\bibnamefont {Timbie}},\
  }\bibfield  {title} {\enquote {\bibinfo {title} {{Calibration of
  millimeter-wave polarimeters using a thin dielectric sheet}},}\ }\href
  {\doibase 10.1109/TMTT.2002.802326} {\bibfield  {journal} {\bibinfo
  {journal} {IEEE Transactions on Microwave Theory and Techniques}\ }\textbf
  {\bibinfo {volume} {50}},\ \bibinfo {pages} {2135--2141} (\bibinfo {year}
  {2002})}\BibitemShut {NoStop}%
\bibitem [{\citenamefont {Weiland}\ \emph {et~al.}(2011)\citenamefont
  {Weiland}, \citenamefont {Odegard}, \citenamefont {Hill} \emph
  {et~al.}}]{Weiland2011}%
  \BibitemOpen
  \bibfield  {author} {\bibinfo {author} {\bibfnamefont {J.~L.}\ \bibnamefont
  {Weiland}}, \bibinfo {author} {\bibfnamefont {N.}~\bibnamefont {Odegard}},
  \bibinfo {author} {\bibfnamefont {R.~S.}\ \bibnamefont {Hill}},  \emph
  {et~al.},\ }\bibfield  {title} {\enquote {\bibinfo {title} {{SEVEN-YEAR
  WILKINSON MICROWAVE ANISOTROPY PROBE (WMAP*) OBSERVATIONS: PLANETS AND
  CELESTIAL CALIBRATION SOURCES}},}\ }\href {\doibase
  10.1088/0067-0049/192/2/19} {\bibfield  {journal} {\bibinfo  {journal} {The
  Astrophysical Journal Supplement Series}\ }\textbf {\bibinfo {volume}
  {192}},\ \bibinfo {pages} {19} (\bibinfo {year} {2011})}\BibitemShut
  {NoStop}%
\bibitem [{\citenamefont {{Aumont, J.}}\ \emph {et~al.}(2010)\citenamefont
  {{Aumont, J.}}, \citenamefont {{Conversi, L.}}, \citenamefont {{Thum, C.}}
  \emph {et~al.}}]{Aumont2010}%
  \BibitemOpen
  \bibfield  {author} {\bibinfo {author} {\bibnamefont {{Aumont, J.}}},
  \bibinfo {author} {\bibnamefont {{Conversi, L.}}}, \bibinfo {author}
  {\bibnamefont {{Thum, C.}}},  \emph {et~al.},\ }\bibfield  {title} {\enquote
  {\bibinfo {title} {{Measurement of the Crab nebula polarization at 90~GHz as
  a calibrator for CMB experiments}},}\ }\href {\doibase
  10.1051/0004-6361/200913834} {\bibfield  {journal} {\bibinfo  {journal}
  {A\&A}\ }\textbf {\bibinfo {volume} {514}},\ \bibinfo {pages} {A70} (\bibinfo
  {year} {2010})}\BibitemShut {NoStop}%
\bibitem [{\citenamefont {{Planck Collaboration}}(2016)}]{Planck2016}%
  \BibitemOpen
  \bibfield  {author} {\bibinfo {author} {\bibnamefont {{Planck
  Collaboration}}},\ }\bibfield  {title} {\enquote {\bibinfo {title} {{Planck
  2015 results - XXVI. The Second Planck Catalogue of Compact Sources}},}\
  }\href {\doibase 10.1051/0004-6361/201526914} {\bibfield  {journal} {\bibinfo
   {journal} {A\&A}\ }\textbf {\bibinfo {volume} {594}},\ \bibinfo {pages}
  {A26} (\bibinfo {year} {2016})}\BibitemShut {NoStop}%
\bibitem [{\citenamefont {{Aumont, J.}}\ \emph {et~al.}(2020)\citenamefont
  {{Aumont, J.}}, \citenamefont {{Mac\'{\i}as-P\'erez, J. F.}}, \citenamefont
  {{Ritacco, A.}} \emph {et~al.}}]{Aumont2020}%
  \BibitemOpen
  \bibfield  {author} {\bibinfo {author} {\bibnamefont {{Aumont, J.}}},
  \bibinfo {author} {\bibnamefont {{Mac\'{\i}as-P\'erez, J. F.}}}, \bibinfo
  {author} {\bibnamefont {{Ritacco, A.}}},  \emph {et~al.},\ }\bibfield
  {title} {\enquote {\bibinfo {title} {{Absolute calibration of the
  polarisation angle for future CMB B-mode experiments from current and future
  measurements of the Crab nebula}},}\ }\href {\doibase
  10.1051/0004-6361/201833504} {\bibfield  {journal} {\bibinfo  {journal}
  {A\&A}\ }\textbf {\bibinfo {volume} {634}},\ \bibinfo {pages} {A100}
  (\bibinfo {year} {2020})}\BibitemShut {NoStop}%
\bibitem [{\citenamefont {Bischoff}\ \emph {et~al.}(2008)\citenamefont
  {Bischoff}, \citenamefont {Hyatt}, \citenamefont {McMahon} \emph
  {et~al.}}]{Bischoff2008}%
  \BibitemOpen
  \bibfield  {author} {\bibinfo {author} {\bibfnamefont {C.}~\bibnamefont
  {Bischoff}}, \bibinfo {author} {\bibfnamefont {L.}~\bibnamefont {Hyatt}},
  \bibinfo {author} {\bibfnamefont {J.~J.}\ \bibnamefont {McMahon}},  \emph
  {et~al.},\ }\bibfield  {title} {\enquote {\bibinfo {title} {{New Measurements
  of Fine-Scale CMB Polarization Power Spectra from CAPMAP at Both 40 and 90
  GHz}},}\ }\href {\doibase 10.1086/590487} {\bibfield  {journal} {\bibinfo
  {journal} {The Astrophysical Journal}\ }\textbf {\bibinfo {volume} {684}},\
  \bibinfo {pages} {771} (\bibinfo {year} {2008})}\BibitemShut {NoStop}%
\bibitem [{\citenamefont {Fedderke}, \citenamefont {Graham},\ and\
  \citenamefont {Rajendran}(2019)}]{Fedderke2019}%
  \BibitemOpen
  \bibfield  {author} {\bibinfo {author} {\bibfnamefont {M.~A.}\ \bibnamefont
  {Fedderke}}, \bibinfo {author} {\bibfnamefont {P.~W.}\ \bibnamefont
  {Graham}}, \ and\ \bibinfo {author} {\bibfnamefont {S.}~\bibnamefont
  {Rajendran}},\ }\bibfield  {title} {\enquote {\bibinfo {title} {{Axion dark
  matter detection with CMB polarization}},}\ }\href {\doibase
  10.1103/PhysRevD.100.015040} {\bibfield  {journal} {\bibinfo  {journal}
  {Phys. Rev. D}\ }\textbf {\bibinfo {volume} {100}},\ \bibinfo {pages}
  {015040} (\bibinfo {year} {2019})}\BibitemShut {NoStop}%
\bibitem [{\citenamefont {Galitzki}\ \emph {et~al.}(2018)\citenamefont
  {Galitzki}, \citenamefont {Ali}, \citenamefont {Arnold} \emph
  {et~al.}}]{Galitzki2018}%
  \BibitemOpen
  \bibfield  {author} {\bibinfo {author} {\bibfnamefont {N.}~\bibnamefont
  {Galitzki}}, \bibinfo {author} {\bibfnamefont {A.}~\bibnamefont {Ali}},
  \bibinfo {author} {\bibfnamefont {K.~S.}\ \bibnamefont {Arnold}},  \emph
  {et~al.},\ }\bibfield  {title} {\enquote {\bibinfo {title} {{The Simons
  Observatory: instrument overview}},}\ }in\ \href {\doibase
  10.1117/12.2312985} {\emph {\bibinfo {booktitle} {Millimeter, Submillimeter,
  and Far-Infrared Detectors and Instrumentation for Astronomy IX}}},\ Vol.\
  \bibinfo {volume} {10708},\ \bibinfo {editor} {edited by\ \bibinfo {editor}
  {\bibfnamefont {J.}~\bibnamefont {Zmuidzinas}}\ and\ \bibinfo {editor}
  {\bibfnamefont {J.-R.}\ \bibnamefont {Gao}}},\ \bibinfo {organization}
  {International Society for Optics and Photonics}\ (\bibinfo  {publisher}
  {SPIE},\ \bibinfo {year} {2018})\ p.\ \bibinfo {pages} {1070804}\BibitemShut
  {NoStop}%
\bibitem [{\citenamefont {Ali}\ \emph {et~al.}(2020)\citenamefont {Ali},
  \citenamefont {Adachi}, \citenamefont {Arnold} \emph {et~al.}}]{Ali2020}%
  \BibitemOpen
  \bibfield  {author} {\bibinfo {author} {\bibfnamefont {A.~M.}\ \bibnamefont
  {Ali}}, \bibinfo {author} {\bibfnamefont {S.}~\bibnamefont {Adachi}},
  \bibinfo {author} {\bibfnamefont {K.}~\bibnamefont {Arnold}},  \emph
  {et~al.},\ }\bibfield  {title} {\enquote {\bibinfo {title} {{Small Aperture
  Telescopes for the Simons Observatory}},}\ }\href {\doibase
  10.1007/s10909-020-02430-5} {\bibfield  {journal} {\bibinfo  {journal}
  {Journal of Low Temperature Physics}\ }\textbf {\bibinfo {volume} {200}},\
  \bibinfo {pages} {461--471} (\bibinfo {year} {2020})}\BibitemShut {NoStop}%
\bibitem [{\citenamefont {Kiuchi}\ \emph {et~al.}(2020)\citenamefont {Kiuchi},
  \citenamefont {Adachi}, \citenamefont {Ali} \emph {et~al.}}]{Kiuchi2020}%
  \BibitemOpen
  \bibfield  {author} {\bibinfo {author} {\bibfnamefont {K.}~\bibnamefont
  {Kiuchi}}, \bibinfo {author} {\bibfnamefont {S.}~\bibnamefont {Adachi}},
  \bibinfo {author} {\bibfnamefont {A.~M.}\ \bibnamefont {Ali}},  \emph
  {et~al.},\ }\bibfield  {title} {\enquote {\bibinfo {title} {{Simons
  Observatory Small Aperture Telescope overview}},}\ }in\ \href {\doibase
  10.1117/12.2562016} {\emph {\bibinfo {booktitle} {Ground-based and Airborne
  Telescopes VIII}}},\ Vol.\ \bibinfo {volume} {11445},\ \bibinfo {editor}
  {edited by\ \bibinfo {editor} {\bibfnamefont {H.~K.}\ \bibnamefont
  {Marshall}}, \bibinfo {editor} {\bibfnamefont {J.}~\bibnamefont
  {Spyromilio}}, \ and\ \bibinfo {editor} {\bibfnamefont {T.}~\bibnamefont
  {Usuda}}},\ \bibinfo {organization} {International Society for Optics and
  Photonics}\ (\bibinfo  {publisher} {SPIE},\ \bibinfo {year} {2020})\ p.\
  \bibinfo {pages} {114457L}\BibitemShut {NoStop}%
\bibitem [{\citenamefont {Galitzki}()}]{Galitzki2023}%
  \BibitemOpen
  \bibfield  {author} {\bibinfo {author} {\bibfnamefont {N.}~\bibnamefont
  {Galitzki}},\ }\href@noop {} {}\bibinfo {howpublished} {"The Simons
  Observatory: Integration and testing of the first small aperture telescope,
  SAT-MF1", in prep}\BibitemShut {NoStop}%
\bibitem [{\citenamefont {Parshley}\ \emph {et~al.}(2018)\citenamefont
  {Parshley}, \citenamefont {Niemack}, \citenamefont {Hills} \emph
  {et~al.}}]{Parshley2018}%
  \BibitemOpen
  \bibfield  {author} {\bibinfo {author} {\bibfnamefont {S.~C.}\ \bibnamefont
  {Parshley}}, \bibinfo {author} {\bibfnamefont {M.}~\bibnamefont {Niemack}},
  \bibinfo {author} {\bibfnamefont {R.}~\bibnamefont {Hills}},  \emph
  {et~al.},\ }\bibfield  {title} {\enquote {\bibinfo {title} {{The optical
  design of the six-meter CCAT-prime and Simons Observatory telescopes}},}\
  }in\ \href {\doibase 10.1117/12.2314073} {\emph {\bibinfo {booktitle}
  {Ground-based and Airborne Telescopes VII}}},\ Vol.\ \bibinfo {volume}
  {10700},\ \bibinfo {editor} {edited by\ \bibinfo {editor} {\bibfnamefont
  {H.~K.}\ \bibnamefont {Marshall}}\ and\ \bibinfo {editor} {\bibfnamefont
  {J.}~\bibnamefont {Spyromilio}}},\ \bibinfo {organization} {International
  Society for Optics and Photonics}\ (\bibinfo  {publisher} {SPIE},\ \bibinfo
  {year} {2018})\ p.\ \bibinfo {pages} {1070041}\BibitemShut {NoStop}%
\bibitem [{\citenamefont {Zhu}\ \emph {et~al.}(2021)\citenamefont {Zhu},
  \citenamefont {Bhandarkar}, \citenamefont {Coppi} \emph {et~al.}}]{Zhu2021}%
  \BibitemOpen
  \bibfield  {author} {\bibinfo {author} {\bibfnamefont {N.}~\bibnamefont
  {Zhu}}, \bibinfo {author} {\bibfnamefont {T.}~\bibnamefont {Bhandarkar}},
  \bibinfo {author} {\bibfnamefont {G.}~\bibnamefont {Coppi}},  \emph
  {et~al.},\ }\bibfield  {title} {\enquote {\bibinfo {title} {{The Simons
  Observatory Large Aperture Telescope Receiver}},}\ }\href {\doibase
  10.3847/1538-4365/ac0db7} {\bibfield  {journal} {\bibinfo  {journal} {The
  Astrophysical Journal Supplement Series}\ }\textbf {\bibinfo {volume}
  {256}},\ \bibinfo {pages} {23} (\bibinfo {year} {2021})}\BibitemShut
  {NoStop}%
\bibitem [{\citenamefont {Irwin}\ and\ \citenamefont
  {Hilton}(2005)}]{Irwin2005}%
  \BibitemOpen
  \bibfield  {author} {\bibinfo {author} {\bibfnamefont {K.}~\bibnamefont
  {Irwin}}\ and\ \bibinfo {author} {\bibfnamefont {G.}~\bibnamefont {Hilton}},\
  }\enquote {\bibinfo {title} {{Transition-Edge Sensors}},}\ in\ \href
  {\doibase 10.1007/10933596_3} {\emph {\bibinfo {booktitle} {Cryogenic
  Particle Detection}}},\ \bibinfo {editor} {edited by\ \bibinfo {editor}
  {\bibfnamefont {C.}~\bibnamefont {Enss}}}\ (\bibinfo  {publisher} {Springer
  Berlin Heidelberg},\ \bibinfo {address} {Berlin, Heidelberg},\ \bibinfo
  {year} {2005})\ pp.\ \bibinfo {pages} {63--150}\BibitemShut {NoStop}%
\bibitem [{\citenamefont {McCarrick}\ \emph {et~al.}(2021)\citenamefont
  {McCarrick}, \citenamefont {Healy}, \citenamefont {Ahmed} \emph
  {et~al.}}]{McCarrick2021}%
  \BibitemOpen
  \bibfield  {author} {\bibinfo {author} {\bibfnamefont {H.}~\bibnamefont
  {McCarrick}}, \bibinfo {author} {\bibfnamefont {E.}~\bibnamefont {Healy}},
  \bibinfo {author} {\bibfnamefont {Z.}~\bibnamefont {Ahmed}},  \emph
  {et~al.},\ }\bibfield  {title} {\enquote {\bibinfo {title} {{The Simons
  Observatory Microwave SQUID Multiplexing Detector Module Design}},}\ }\href
  {\doibase 10.3847/1538-4357/ac2232} {\bibfield  {journal} {\bibinfo
  {journal} {The Astrophysical Journal}\ }\textbf {\bibinfo {volume} {922}},\
  \bibinfo {pages} {38} (\bibinfo {year} {2021})}\BibitemShut {NoStop}%
\bibitem [{\citenamefont {Kusaka}\ \emph {et~al.}(2014)\citenamefont {Kusaka},
  \citenamefont {Essinger-Hileman}, \citenamefont {Appel} \emph
  {et~al.}}]{Kusaka2014}%
  \BibitemOpen
  \bibfield  {author} {\bibinfo {author} {\bibfnamefont {A.}~\bibnamefont
  {Kusaka}}, \bibinfo {author} {\bibfnamefont {T.}~\bibnamefont
  {Essinger-Hileman}}, \bibinfo {author} {\bibfnamefont {J.~W.}\ \bibnamefont
  {Appel}},  \emph {et~al.},\ }\bibfield  {title} {\enquote {\bibinfo {title}
  {{Modulation of cosmic microwave background polarization with a warm rapidly
  rotating half-wave plate on the Atacama B-Mode Search instrument}},}\ }\href
  {\doibase 10.1063/1.4862058} {\bibfield  {journal} {\bibinfo  {journal}
  {Review of Scientific Instruments}\ }\textbf {\bibinfo {volume} {85}}
  (\bibinfo {year} {2014}),\ 10.1063/1.4862058},\ \bibinfo {note}
  {024501}\BibitemShut {NoStop}%
\bibitem [{\citenamefont {Takakura}\ \emph {et~al.}(2017)\citenamefont
  {Takakura}, \citenamefont {Aguilar}, \citenamefont {Akiba} \emph
  {et~al.}}]{Takakura2017}%
  \BibitemOpen
  \bibfield  {author} {\bibinfo {author} {\bibfnamefont {S.}~\bibnamefont
  {Takakura}}, \bibinfo {author} {\bibfnamefont {M.}~\bibnamefont {Aguilar}},
  \bibinfo {author} {\bibfnamefont {Y.}~\bibnamefont {Akiba}},  \emph
  {et~al.},\ }\bibfield  {title} {\enquote {\bibinfo {title} {{Performance of a
  continuously rotating half-wave plate on the POLARBEAR telescope}},}\ }\href
  {\doibase 10.1088/1475-7516/2017/05/008} {\bibfield  {journal} {\bibinfo
  {journal} {Journal of Cosmology and Astroparticle Physics}\ }\textbf
  {\bibinfo {volume} {2017}},\ \bibinfo {pages} {008} (\bibinfo {year}
  {2017})}\BibitemShut {NoStop}%
\bibitem [{\citenamefont {Yamada}()}]{Yamada2022}%
  \BibitemOpen
  \bibfield  {author} {\bibinfo {author} {\bibfnamefont {K.}~\bibnamefont
  {Yamada}},\ }\href@noop {} {}\bibinfo {howpublished} {"The Simons
  Observatory: Cryogenic Half-Wave Plate Rotation Mechanism for the Small
  Aperture Telescope", in prep}\BibitemShut {NoStop}%
\bibitem [{\citenamefont {Coppi}\ \emph {et~al.}(2022)\citenamefont {Coppi},
  \citenamefont {Conenna}, \citenamefont {Savorgnano} \emph
  {et~al.}}]{Coppi2022}%
  \BibitemOpen
  \bibfield  {author} {\bibinfo {author} {\bibfnamefont {G.}~\bibnamefont
  {Coppi}}, \bibinfo {author} {\bibfnamefont {G.}~\bibnamefont {Conenna}},
  \bibinfo {author} {\bibfnamefont {S.}~\bibnamefont {Savorgnano}},  \emph
  {et~al.},\ }\bibfield  {title} {\enquote {\bibinfo {title} {{PROTOCALC: an
  artificial calibrator source for CMB telescopes}},}\ }in\ \href {\doibase
  10.1117/12.2628312} {\emph {\bibinfo {booktitle} {Millimeter, Submillimeter,
  and Far-Infrared Detectors and Instrumentation for Astronomy XI}}},\ Vol.\
  \bibinfo {volume} {12190},\ \bibinfo {editor} {edited by\ \bibinfo {editor}
  {\bibfnamefont {J.}~\bibnamefont {Zmuidzinas}}\ and\ \bibinfo {editor}
  {\bibfnamefont {J.-R.}\ \bibnamefont {Gao}}},\ \bibinfo {organization}
  {International Society for Optics and Photonics}\ (\bibinfo  {publisher}
  {SPIE},\ \bibinfo {year} {2022})\ p.\ \bibinfo {pages} {1219015}\BibitemShut
  {NoStop}%
\bibitem [{\citenamefont {Tajima}\ \emph {et~al.}(2012)\citenamefont {Tajima},
  \citenamefont {Nguyen}, \citenamefont {Bischoff} \emph
  {et~al.}}]{Tajima2012}%
  \BibitemOpen
  \bibfield  {author} {\bibinfo {author} {\bibfnamefont {O.}~\bibnamefont
  {Tajima}}, \bibinfo {author} {\bibfnamefont {H.}~\bibnamefont {Nguyen}},
  \bibinfo {author} {\bibfnamefont {C.}~\bibnamefont {Bischoff}},  \emph
  {et~al.},\ }\bibfield  {title} {\enquote {\bibinfo {title} {Novel calibration
  system with sparse wires for cmb polarization receivers},}\ }\href {\doibase
  10.1007/s10909-012-0545-3} {\bibfield  {journal} {\bibinfo  {journal}
  {Journal of Low Temperature Physics}\ }\textbf {\bibinfo {volume} {167}},\
  \bibinfo {pages} {936--942} (\bibinfo {year} {2012})}\BibitemShut {NoStop}%
\bibitem [{\citenamefont {Simon}\ \emph {et~al.}(2014)\citenamefont {Simon},
  \citenamefont {Appel}, \citenamefont {Cho} \emph {et~al.}}]{Simon2014}%
  \BibitemOpen
  \bibfield  {author} {\bibinfo {author} {\bibfnamefont {S.~M.}\ \bibnamefont
  {Simon}}, \bibinfo {author} {\bibfnamefont {J.~W.}\ \bibnamefont {Appel}},
  \bibinfo {author} {\bibfnamefont {H.~M.}\ \bibnamefont {Cho}},  \emph
  {et~al.},\ }\bibfield  {title} {\enquote {\bibinfo {title} {{In Situ Time
  Constant and Optical Efficiency Measurements of TRUCE Pixels in the Atacama
  B-Mode Search}},}\ }\href {\doibase 10.1007/s10909-013-0999-y} {\bibfield
  {journal} {\bibinfo  {journal} {Journal of Low Temperature Physics}\ }\textbf
  {\bibinfo {volume} {176}},\ \bibinfo {pages} {712--718} (\bibinfo {year}
  {2014})}\BibitemShut {NoStop}%
\bibitem [{\citenamefont {Kusaka}\ \emph {et~al.}(2018)\citenamefont {Kusaka},
  \citenamefont {Appel}, \citenamefont {Essinger-Hileman} \emph
  {et~al.}}]{Kusaka2018}%
  \BibitemOpen
  \bibfield  {author} {\bibinfo {author} {\bibfnamefont {A.}~\bibnamefont
  {Kusaka}}, \bibinfo {author} {\bibfnamefont {J.}~\bibnamefont {Appel}},
  \bibinfo {author} {\bibfnamefont {T.}~\bibnamefont {Essinger-Hileman}},
  \emph {et~al.},\ }\bibfield  {title} {\enquote {\bibinfo {title} {{Results
  from the Atacama B-mode Search (ABS) experiment}},}\ }\href {\doibase
  10.1088/1475-7516/2018/09/005} {\bibfield  {journal} {\bibinfo  {journal}
  {Journal of Cosmology and Astroparticle Physics}\ }\textbf {\bibinfo {volume}
  {2018}},\ \bibinfo {pages} {005} (\bibinfo {year} {2018})}\BibitemShut
  {NoStop}%
\bibitem [{\citenamefont {Chinone}(2010)}]{Chinone2010}%
  \BibitemOpen
  \bibfield  {author} {\bibinfo {author} {\bibfnamefont {Y.}~\bibnamefont
  {Chinone}},\ }\bibfield  {title} {\enquote {\bibinfo {title} {{Measurement of
  cosmic microwave background polarization power spectra at 43 GHz with Q/U
  Imaging ExperimenT}},}\ }\href@noop {} {\bibfield  {journal} {\bibinfo
  {journal} {Ddissertation for Doctoral Degree. Sendai-shi: Tohoku University}\
  } (\bibinfo {year} {2010})}\BibitemShut {NoStop}%
\bibitem [{\citenamefont {Matsuda}()}]{Matsuda2022}%
  \BibitemOpen
  \bibfield  {author} {\bibinfo {author} {\bibfnamefont {F.}~\bibnamefont
  {Matsuda}},\ }\href@noop {} {}\bibinfo {howpublished} {"Optics Design of the
  Simons Observatory Small Aperture Telescopes", in prep}\BibitemShut {NoStop}%
\bibitem [{\citenamefont {{E$\&$C Engineering}}(2022)}]{EC}%
  \BibitemOpen
  \bibfield  {author} {\bibinfo {author} {\bibnamefont {{E$\&$C
  Engineering}}},\ }\href@noop {} {}\bibinfo {howpublished}
  {\url{https://ece.co.jp/en/products/absorber/an.html}} (\bibinfo {year}
  {2022})\BibitemShut {NoStop}%
\bibitem [{\citenamefont {Weast}, \citenamefont {Astle},\ and\ \citenamefont
  {Beyer}(2011)}]{Weast2011}%
  \BibitemOpen
  \bibfield  {author} {\bibinfo {author} {\bibfnamefont {R.~C.}\ \bibnamefont
  {Weast}}, \bibinfo {author} {\bibfnamefont {M.~J.}\ \bibnamefont {Astle}}, \
  and\ \bibinfo {author} {\bibfnamefont {W.~H.}\ \bibnamefont {Beyer}},\
  }\href@noop {} {\emph {\bibinfo {title} {{CRC handbook of chemistry and
  physics}}}},\ Vol.\ \bibinfo {volume} {91th ed. : 2010-2011}\ (\bibinfo
  {publisher} {Boca Raton, Fla. : CRC Press},\ \bibinfo {year}
  {2011})\BibitemShut {NoStop}%
\bibitem [{\citenamefont {Rieck}(1967)}]{Rieck1967}%
  \BibitemOpen
  \bibfield  {author} {\bibinfo {author} {\bibfnamefont {G.~D.}\ \bibnamefont
  {Rieck}},\ }\href@noop {} {\emph {\bibinfo {title} {{Tungsten and its
  compounds}}}}\ (\bibinfo  {publisher} {Oxford ; New York : Pergamon Press},\
  \bibinfo {year} {1967})\BibitemShut {NoStop}%
\bibitem [{\citenamefont {{Silverthin}}(2022)}]{Silverthin}%
  \BibitemOpen
  \bibfield  {author} {\bibinfo {author} {\bibnamefont {{Silverthin}}},\
  }\href@noop {} {}\bibinfo {howpublished} {\url{https://www.silverthin.com/}}
  (\bibinfo {year} {2022})\BibitemShut {NoStop}%
\bibitem [{\citenamefont {{Renishaw}}(2022)}]{Renishaw}%
  \BibitemOpen
  \bibfield  {author} {\bibinfo {author} {\bibnamefont {{Renishaw}}},\
  }\href@noop {} {}\bibinfo {howpublished}
  {\url{https://www.rls.si/eng/lm15-linear-and-rotary-magnetic-encoder-system}}
  (\bibinfo {year} {2022})\BibitemShut {NoStop}%
\bibitem [{\citenamefont {{Digi-pas}}(2022)}]{Digi-pas}%
  \BibitemOpen
  \bibfield  {author} {\bibinfo {author} {\bibnamefont {{Digi-pas}}},\
  }\href@noop {} {}\bibinfo {howpublished} {\url{https://www.digipas.com/}}
  (\bibinfo {year} {2022})\BibitemShut {NoStop}%
\bibitem [{\citenamefont {{OpenBuilds}}(2022)}]{OpenBuilds}%
  \BibitemOpen
  \bibfield  {author} {\bibinfo {author} {\bibnamefont {{OpenBuilds}}},\
  }\href@noop {} {}\bibinfo {howpublished}
  {\url{https://openbuildspartstore.com/v-slot-nema-23-linear-actuator-belt-driven/}}
  (\bibinfo {year} {2022})\BibitemShut {NoStop}%
\bibitem [{\citenamefont {{Orientalmotor}}(2022)}]{Orientalmotor}%
  \BibitemOpen
  \bibfield  {author} {\bibinfo {author} {\bibnamefont {{Orientalmotor}}},\
  }\href@noop {} {}\bibinfo {howpublished}
  {\url{https://www.orientalmotor.com/}} (\bibinfo {year} {2022})\BibitemShut
  {NoStop}%
\bibitem [{\citenamefont {{Omron}}(2023)}]{Omron}%
  \BibitemOpen
  \bibfield  {author} {\bibinfo {author} {\bibnamefont {{Omron}}},\ }\href@noop
  {} {}\bibinfo {howpublished}
  {\url{https://www.fa.omron.co.jp/product/item/8235/}} (\bibinfo {year}
  {2023})\BibitemShut {NoStop}%
\bibitem [{\citenamefont {{Laird}}(2022)}]{Laird}%
  \BibitemOpen
  \bibfield  {author} {\bibinfo {author} {\bibnamefont {{Laird}}},\ }\href@noop
  {} {}\bibinfo {howpublished} {\url{https://www.lairdtech.com/}} (\bibinfo
  {year} {2022})\BibitemShut {NoStop}%
\bibitem [{\citenamefont {{Ohnishi}}(2022)}]{Ohnishi}%
  \BibitemOpen
  \bibfield  {author} {\bibinfo {author} {\bibnamefont {{Ohnishi}}},\
  }\href@noop {} {}\bibinfo {howpublished} {\url{https://www.oss-ohnishi.com/}}
  (\bibinfo {year} {2022})\BibitemShut {NoStop}%
\bibitem [{\citenamefont {{Faro}}(2022)}]{Faro}%
  \BibitemOpen
  \bibfield  {author} {\bibinfo {author} {\bibnamefont {{Faro}}},\ }\href@noop
  {} {}\bibinfo {howpublished}
  {\url{https://www.faro.com/en/Products/Hardware/Quantum-FaroArms}} (\bibinfo
  {year} {2022})\BibitemShut {NoStop}%
\bibitem [{\citenamefont {Osipov}\ and\ \citenamefont
  {Tretyakov}(2017)}]{Osipov2017}%
  \BibitemOpen
  \bibfield  {author} {\bibinfo {author} {\bibfnamefont {A.~V.}\ \bibnamefont
  {Osipov}}\ and\ \bibinfo {author} {\bibfnamefont {S.~A.}\ \bibnamefont
  {Tretyakov}},\ }\href@noop {} {\emph {\bibinfo {title} {{Modern
  electromagnetic scattering theory with applications}}}}\ (\bibinfo
  {publisher} {John Wiley \& Sons},\ \bibinfo {year} {2017})\BibitemShut
  {NoStop}%
\end{thebibliography}%




\end{document}